\pdfoutput=1 
\documentclass[cits]{JINST}

\usepackage{graphicx}
\usepackage{amssymb}
\usepackage{amsthm}
\usepackage{mathtools}
\usepackage{tabularx}
\usepackage{textcomp}
\usepackage{cite}
\usepackage[english]{babel}
\usepackage{booktabs}

\title{Nanosecond-level time synchronization of autonomous radio detector stations
for extensive air showers}

\author{
The Pierre Auger Collaboration
}\author{
A.~Aab$^{41}$,
P.~Abreu$^{67}$,
M.~Aglietta$^{52,53}$,
E.J.~Ahn$^{82}$,
I.~Al Samarai$^{28}$,
I.F.M.~Albuquerque$^{16}$,
I.~Allekotte$^{1}$,
P.~Allison$^{87}$,
A.~Almela$^{11,8}$,
J.~Alvarez Castillo$^{60}$
% ...

\vskip 0.5cm
\normalfont
\begin{flushleft}
\begin{itshape}
\begin{small}
\par\noindent
{$^{1}$} Centro At\'omico Bariloche and Instituto Balseiro (CNEA-UNCuyo-CONICET), San Carlos de Bariloche, Argentina\\
{$^{2}$} Centro de Investigaciones en L\'aseres y Aplicaciones, CITEDEF and CONICET, Villa Martelli, Argentina\\
{$^{3}$} Departamento de F\'\i{}sica, FCEyN, Universidad de Buenos Aires and CONICET, Buenos Aires, Argentina\\
{$^{4}$} IFLP, Universidad Nacional de La Plata and CONICET, La Plata, Argentina\\
{$^{5}$} Instituto de Astronom\'\i{}a y F\'\i{}sica del Espacio (IAFE, CONICET-UBA), Buenos Aires, Argentina\\
{$^{6}$} Instituto de F\'\i{}sica de Rosario (IFIR) -- CONICET/U.N.R.\ and Facultad de Ciencias Bioqu\'\i{}micas y Farmac\'euticas U.N.R., Rosario, Argentina\\
{$^{7}$} Instituto de Tecnolog\'\i{}as en Detecci\'on y Astropart\'\i{}culas (CNEA, CONICET, UNSAM), and Universidad Tecnol\'ogica Nacional -- Facultad Regional Mendoza (CONICET/CNEA), Mendoza, Argentina\\
{$^{8}$} Instituto de Tecnolog\'\i{}as en Detecci\'on y Astropart\'\i{}culas (CNEA, CONICET, UNSAM), Buenos Aires, Argentina\\
{$^{9}$} Observatorio Pierre Auger, Malarg\"ue, Argentina\\
{$^{10}$} Observatorio Pierre Auger and Comisi\'on Nacional de Energ\'\i{}a At\'omica, Malarg\"ue, Argentina\\
{$^{11}$} Universidad Tecnol\'ogica Nacional -- Facultad Regional Buenos Aires, Buenos Aires, Argentina\\
{$^{12}$} University of Adelaide, Adelaide, S.A., Australia\\
{$^{13}$} Centro Brasileiro de Pesquisas Fisicas, Rio de Janeiro, RJ, Brazil\\
{$^{14}$} Universidade de S\~ao Paulo, Escola de Engenharia de Lorena, Lorena, SP, Brazil\\
{$^{15}$} Universidade de S\~ao Paulo, Instituto de F\'\i{}sica de S\~ao Carlos, S\~ao Carlos, SP, Brazil\\
{$^{16}$} Universidade de S\~ao Paulo, Instituto de F\'\i{}sica, S\~ao Paulo, SP, Brazil\\
{$^{17}$} Universidade Estadual de Campinas, IFGW, Campinas, SP, Brazil\\
{$^{18}$} Universidade Estadual de Feira de Santana, Feira de Santana, Brazil\\
{$^{19}$} Universidade Federal da Bahia, Salvador, BA, Brazil\\
{$^{20}$} Universidade Federal de Pelotas, Pelotas, RS, Brazil\\
{$^{21}$} Universidade Federal do ABC, Santo Andr\'e, SP, Brazil\\
{$^{22}$} Universidade Federal do Rio de Janeiro, Instituto de F\'\i{}sica, Rio de Janeiro, RJ, Brazil\\
{$^{23}$} Universidade Federal Fluminense, EEIMVR, Volta Redonda, RJ, Brazil\\
{$^{24}$} Universidad Industrial de Santander, Bucaramanga, Colombia\\
{$^{25}$} Charles University, Faculty of Mathematics and Physics, Institute of Particle and Nuclear Physics, Prague, Czech Republic\\
{$^{26}$} Institute of Physics of the Academy of Sciences of the Czech Republic, Prague, Czech Republic\\
{$^{27}$} Palacky University, RCPTM, Olomouc, Czech Republic\\
{$^{28}$} Institut de Physique Nucl\'eaire d'Orsay (IPNO), Universit\'e Paris 11, CNRS-IN2P3, Orsay, France\\
{$^{29}$} Laboratoire de l'Acc\'el\'erateur Lin\'eaire (LAL), Universit\'e Paris 11, CNRS-IN2P3, Orsay, France\\
{$^{30}$} Laboratoire de Physique Nucl\'eaire et de Hautes Energies (LPNHE), Universit\'es Paris 6 et Paris 7, CNRS-IN2P3, Paris, France\\
{$^{31}$} Laboratoire de Physique Subatomique et de Cosmologie (LPSC), Universit\'e Grenoble-Alpes, CNRS/IN2P3, Grenoble, France\\
{$^{32}$} Station de Radioastronomie de Nan\c{c}ay, Observatoire de Paris, CNRS/INSU, Nan\c{c}ay, France\\
\end{small}
\end{itshape}
\end{flushleft}
} \author{
\normalfont
\begin{flushleft}
\begin{itshape}
\begin{small}
\par\noindent
{$^{33}$} SUBATECH, \'Ecole des Mines de Nantes, CNRS-IN2P3, Universit\'e de Nantes, Nantes, France\\
{$^{34}$} Bergische Universit\"at Wuppertal, Fachbereich C -- Physik, Wuppertal, Germany\\
{$^{35}$} Karlsruhe Institute of Technology (KIT) -- Campus South -- Institut f\"ur Experimentelle Kernphysik (IEKP), Karlsruhe, Germany\\
{$^{36}$} Karlsruhe Institute of Technology (KIT) -- Campus North -- Institut f\"ur Kernphysik (IKP), Karlsruhe, Germany\\
{$^{37}$} Karlsruhe Institute of Technology (KIT) -- Campus North -- Institut f\"ur Prozessdatenverarbeitung und Elektronik (IEKP), Karlsruhe, Germany\\
{$^{38}$} Max-Planck-Institut f\"ur Radioastronomie, Bonn, Germany\\
{$^{39}$} RWTH Aachen University, III.\ Physikalisches Institut A, Aachen, Germany\\
{$^{40}$} Universit\"at Hamburg, II.\ Institut f\"ur Theoretische Physik, Hamburg, Germany\\
{$^{41}$} Universit\"at Siegen, Fachbereich 7 Physik -- Experimentelle Teilchenphysik, Siegen, Germany\\
{$^{42}$} Universit\`a di Milano and Sezione INFN, Milan, Italy\\
{$^{43}$} Universit\`a di Napoli ``Federico II'' and Sezione INFN, Napoli, Italy\\
{$^{44}$} Universit\`a di Roma II ``Tor Vergata'' and Sezione INFN, Roma, Italy\\
{$^{45}$} Universit\`a di Catania and Sezione INFN, Catania, Italy\\
{$^{46}$} Universit\`a di Torino and Sezione INFN, Torino, Italy\\
{$^{47}$} Dipartimento di Matematica e Fisica ``E.\ De Giorgi'' dell'Universit\`a del Salento and Sezione INFN, Lecce, Italy\\
{$^{48}$} Dipartimento di Scienze Fisiche e Chimiche dell'Universit\`a dell'Aquila and Sezione INFN, L'Aquila, Italy\\
{$^{49}$} Gran Sasso Science Institute (INFN), L'Aquila, Italy\\
{$^{50}$} Istituto di Astrofisica Spaziale e Fisica Cosmica di Palermo (INAF), Palermo, Italy\\
{$^{51}$} INFN, Laboratori Nazionali del Gran Sasso, Assergi (L'Aquila), Italy\\
{$^{52}$} Osservatorio Astrofisico di Torino (INAF), Torino, Italy\\
{$^{53}$} INFN, Sezione di Torino, Italy\\
{$^{54}$} Universit\`a di Torino, Torino, Italy\\
{$^{55}$} Benem\'erita Universidad Aut\'onoma de Puebla, Puebla, M\'exico\\
{$^{56}$} Centro de Investigaci\'on y de Estudios Avanzados del IPN (CINVESTAV), M\'exico, D.F., M\'exico\\
{$^{57}$} Unidad Profesional Interdisciplinaria en Ingenier\'\i{}a y Tecnolog\'\i{}as Avanzadas del Instituto Polit\'ecnico Nacional (UPIITA-IPN), M\'exico, D.F., M\'exico\\
{$^{58}$} Universidad Aut\'onoma de Chiapas, Tuxtla Guti\'errez, Chiapas, M\'exico\\
{$^{59}$} Universidad Michoacana de San Nicol\'as de Hidalgo, Morelia, Michoac\'an, M\'exico\\
{$^{60}$} Universidad Nacional Aut\'onoma de M\'exico, M\'exico, D.F., M\'exico\\
{$^{61}$} IMAPP, Radboud University Nijmegen, Nijmegen, Netherlands\\
{$^{62}$} KVI -- Center for Advanced Radiation Technology, University of Groningen, Groningen, Netherlands\\
{$^{63}$} Nikhef, Science Park, Amsterdam, Netherlands\\
{$^{64}$} ASTRON, Dwingeloo, Netherlands\\
{$^{65}$} Institute of Nuclear Physics PAN, Krakow, Poland\\
{$^{66}$} University of \L{}\'od\'z, \L{}\'od\'z, Poland\\
{$^{67}$} Laborat\'orio de Instrumenta\c{c}\~ao e F\'\i{}sica Experimental de Part\'\i{}culas (LIP) and Instituto Superior T\'ecnico, Universidade de Lisboa (UL), Portugal\\
{$^{68}$} ``Horia Hulubei'' National Institute for Physics and Nuclear Engineering, Bucharest-Magurele, Romania\\
{$^{69}$} Institute of Space Science, Bucharest-Magurele, Romania\\
{$^{70}$} University of Bucharest, Physics Department, Bucharest, Romania\\
{$^{71}$} University Politehnica of Bucharest, Bucharest, Romania\\
{$^{72}$} Experimental Particle Physics Department, J.\ Stefan Institute, Ljubljana, Slovenia\\
{$^{73}$} Laboratory for Astroparticle Physics, University of Nova Gorica, Nova Gorica, Slovenia\\
{$^{74}$} Universidad Complutense de Madrid, Madrid, Spain\\
\end{small}
\end{itshape}
\end{flushleft}
} \author{
\normalfont
\begin{flushleft}
\begin{itshape}
\begin{small}
\par\noindent
{$^{75}$} Universidad de Alcal\'a, Alcal\'a de Henares, Madrid, Spain\\
{$^{76}$} Universidad de Granada and C.A.F.P.E., Granada, Spain\\
{$^{77}$} Universidad de Santiago de Compostela, Santiago de Compostela, Spain\\
{$^{78}$} Case Western Reserve University, Cleveland, OH, USA\\
{$^{79}$} Colorado School of Mines, Golden, CO, USA\\
{$^{80}$} Colorado State University, Fort Collins, CO, USA\\
{$^{81}$} Department of Physics and Astronomy, Lehman College, City University of New York, Bronx, NY, USA\\
{$^{82}$} Fermilab, Batavia, IL, USA\\
{$^{83}$} Louisiana State University, Baton Rouge, LA, USA\\
{$^{84}$} Michigan Technological University, Houghton, MI, USA\\
{$^{85}$} New York University, New York, NY, USA\\
{$^{86}$} Northeastern University, Boston, MA, USA\\
{$^{87}$} Ohio State University, Columbus, OH, USA\\
{$^{88}$} Pennsylvania State University, University Park, PA, USA\\
{$^{89}$} University of Chicago, Enrico Fermi Institute, Chicago, IL, USA\\
{$^{90}$} University of Hawaii, Honolulu, HI, USA\\
{$^{91}$} University of Nebraska, Lincoln, NE, USA\\
{$^{92}$} University of New Mexico, Albuquerque, NM, USA\\
{$^{a}$} School of Physics and Astronomy, University of Leeds, Leeds, United Kingdom\\
{$^{b}$} Also at Vrije Universiteit Brussels, Brussels, Belgium
\end{small}
\end{itshape}
\end{flushleft}

  E-mail: \href{mailto:auger_spokespersons@fnal.gov}{\rm auger\_spokespersons@fnal.gov} 
}  

\abstract{
To exploit the full potential of radio measurements of cosmic-ray air 
showers at MHz frequencies, a detector timing synchronization within 1~ns is needed. Large distributed 
radio detector arrays such as the Auger Engineering Radio Array (AERA) 
rely on timing via the Global Positioning System (GPS) for the synchronization of individual detector 
station clocks. Unfortunately, GPS timing is expected to have an accuracy no better than about 5~ns. In practice, in particular in AERA, the GPS 
clocks exhibit drifts on the order of tens of ns. We developed a technique to correct for the GPS drifts, 
and an independent method is used to cross-check that indeed we reach a nanosecond-scale timing accuracy by this correction. First, we 
operate a ``beacon transmitter'' which emits defined sine waves detected by AERA antennas 
recorded within the physics data. The relative phasing of these sine 
waves can be used to correct for GPS clock drifts. In addition to this, we observe radio 
pulses emitted by commercial airplanes, the position of which we 
determine in real time from Automatic Dependent Surveillance Broadcasts intercepted with a 
software-defined radio. From the known source location and the 
measured arrival times of the pulses we determine relative timing 
offsets between radio detector stations. We demonstrate with a combined analysis that 
the two methods give a consistent timing calibration with an accuracy 
of 2~ns or better. Consequently, the beacon method alone 
can be used in the future to continuously determine and correct for GPS clock 
drifts in each individual event measured by AERA.
}


\begin{document}


\section{Introduction}

In the last few years, radio detection of cosmic rays has matured from 
small prototype installations to full-fledged experiments contributing 
valuable information in the field of cosmic-ray physics 
\cite{HuegeReviewIcrc2013}. In particular, radio detection has proven to be 
able to extract information on the energy 
\cite{ApelArteagaBaehren2014, AugerEnergyPRD2015} and mass \cite{AERA_PISA2015, ApelArteagaBaehren2014,LOFARXmaxMethod2014, TunkaRexJCAP2015} of the 
primary cosmic rays with a quality competitive with other detection 
techniques. To maximize the potential of radio detection 
arrays, a precise time synchronization of individual radio detector 
stations is needed. This is true in particular if interferometric 
analysis techniques are to be employed \cite{FalckeNature2005}. Another analysis strategy 
that requires very precise timing is the determination of the opening 
angle of the hyperbolic wavefront emitted by extensive air showers 
to deduce mass-sensitive parameters \cite{LOPESWavefront2014}. The consequence is that 
any radio detectors operating in the frequency band below 100~MHz need a 
time synchronization on the order of 1~ns to exploit their full potential.

\begin{figure*}[t]
  \centering
  \includegraphics[width=0.7\textwidth]{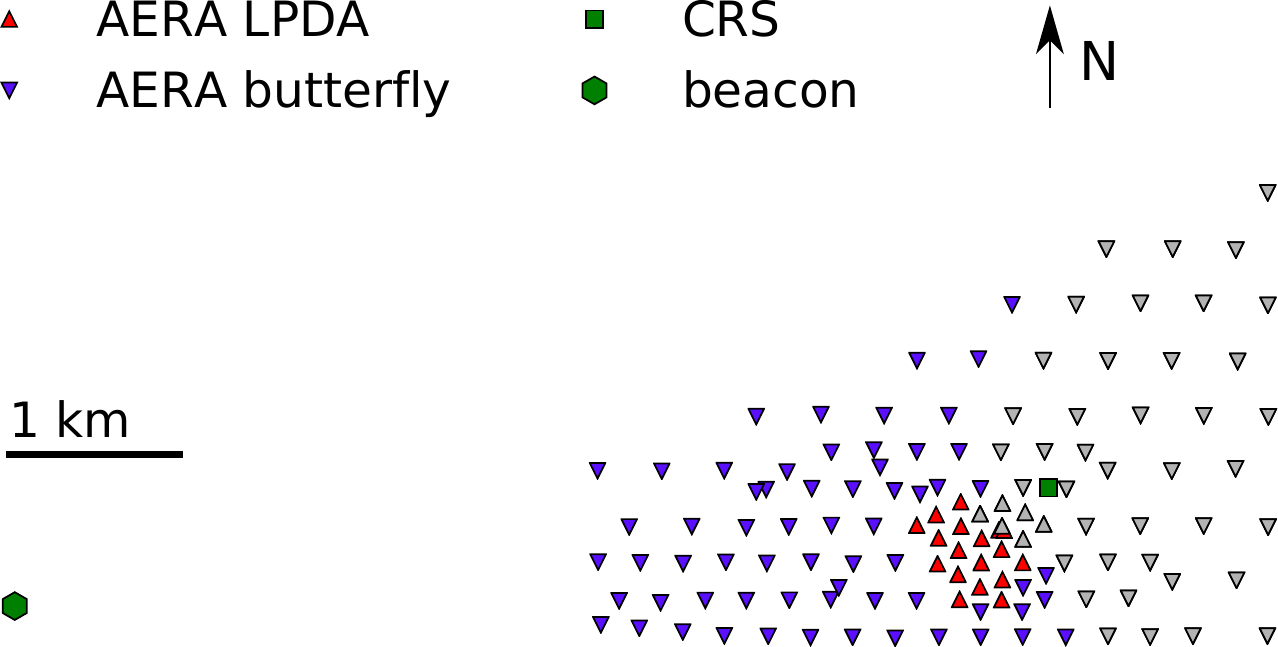}
  \caption{Map of the radio detector stations of the Auger Engineering 
  Radio Array in the year 2014. Detector stations with either butterfly or logarithmic periodic dipole antennas (LPDAs)
  have been deployed \cite{AERAAntennaPaper2012}. The position of the beacon transmitter is west 
  of the array, at the site of the \lq Coihueco\rq~fluorescence telescope building. Only the subset of radio detector stations capable 
  of deep buffering for an external trigger, marked with red and blue 
  triangles, was used for results presented in this paper (other stations are marked in gray). The ADS-B reception hardware 
  is installed at the Central Radio Station (CRS) in the middle of 
  AERA. Not shown here for simplicity: 25 extra antenna stations of butterfly type deployed in the south in 2015; 
  Auger particle detectors at the same site (cf.~Refs.\cite{AERA_PISA2015, AugerDescriptionPaper2015}).}
  \label{fig:CR_aeramap}
\end{figure*}

Nanosecond-scale timing synchronization has previously been achieved in 
cabled setups for radio detection of cosmic-ray air showers 
\cite{SchroederAschBaehren2010,LOFARWavefront2014}. In fact, mature 
open-source solutions such as ``White Rabbit'' \cite{WhiteRabbit} 
have been developed in the past few years and are nowadays used by
a broad community to reliably achieve sub-nanosecond timing synchronization in distributed 
detectors. However, these solutions rely on cabled links to distribute 
timing signals in a deterministic and stable way. 
The time synchronization of autonomous detector 
stations connected only via wireless communications such as those operated 
in AERA \cite{HuegePisa2009} (see Fig. \ref{fig:CR_aeramap}) 
at the Pierre Auger Observatory \cite{AugerDescriptionPaper2015} 
constitute a much more difficult challenge. 

AERA consists of $153$ autonomous stations featuring different spacings of $150\,$m to $750\,$m on an area of $17\,$km\textsuperscript{2}. 
Each station features its individual solar power supply, and local data-acquisition with self-trigger on the radio signal. 
In addition, an external trigger by the Auger particle and fluorescence detectors is sent through the wireless communication to a part of the radio array. 
The same external trigger mechanism is also used to communicate a forced trigger at a rate of $0.01\,$Hz to each station, 
which provides a data set of coincident background measurements. 
For the proof-of-principle of the time synchronization methods presented here, only a part of the array is used, namely only those stations 
which feature an external trigger in addition to the self-trigger, and which have been commissioned before 2015. Later, the methods will be applied to the other stations.  
The used stations are marked with colored triangles in Fig. \ref{fig:CR_aeramap}.

Time synchronization at AERA is performed via a GPS clock in each station, which is in principle capable of reaching an accuracy of $\sim 5$~ns \cite{iLotusGPS}, 
in particular since equal cable lengths are used in all stations for all types of signal cables.
In practice, however, it is often seen that this timing precision is only 
achieved over short time scales and that drifts of up to several tens of
nanoseconds occur over the course of a day. This is true in particular for the 
individual radio detector stations of AERA. GPS-disciplined 
oscillators could possibly improve timing stability on 
scales of minutes to hours, and should be investigated as an option for 
future distributed radio detectors. For AERA, they were not considered 
for various reasons (focus on relative timing, cost issues, 
increased complexity and thus possibly higher rate of failure in the field).

To nevertheless reach the desired nanosecond-scale timing precision, 
alternative measures were employed within AERA. In particular, we 
operate a ``beacon reference transmitter'', originally pioneered in 
the context of the LOPES experiment \cite{SchroederAschBaehren2010} to achieve precise timing on 
an event-by-event basis. In this approach, sine waves at 
several frequencies are continuously emitted by a dedicated 
transmitter and recorded within the data stream of measured cosmic-ray events. Using the relative 
phases with which the different sine waves arrive at each individual 
AERA detector station, possible timing drifts of the station GPS 
clocks can be corrected for. 
However, one has to verify that systematic uncertainties, such as variable timing 
offsets introduced by changing ground and atmospheric conditions during the propagation of the beacon signal from 
the transmitter to the detector stations, are small compared to the desired accuracy of $1\,$ns.

Therefore, an independent second method to perform a nanosecond-scale 
timing calibration of AERA detector stations was developed as a cross-check. It makes 
use of the fact that commercial airplanes emit radio pulses which can 
be seen in the physics data recorded by AERA. Such signals have 
previously been used in prototype setups of AERA to determine 
direction resolution and relative amplitude calibration 
\cite{RevenuAirplane2011}. In combination with real-time position 
information about the emitting airplane, which is available via 
digital ADS-B messages sent by 
the airplane, pulsed airplane signals can be used to 
independently check the time synchronization of the AERA detector 
stations and thus verify the validity of the drift correction performed with the 
beacon transmitter technique.

In this article, we first discuss the beacon transmitter technique in section
\ref{sec:beacon} and the airplane timing analysis technique in section 
\ref{sec:airplane}. Afterwards, we combine the two analyses in section 
\ref{sec:combined} to cross-check them against each other and 
demonstrate that they give consistent results within 2~ns, meaning in particular that the beacon technique is precise to at least 2~ns. We finish with a discussion of remaining limitations of the techniques in section \ref{sec:discussion} 
and our conclusions in section \ref{sec:conclusions}.


\section{The AERA reference beacon} \label{sec:beacon}

In this section we give technical details of the beacon transmitter, 
describe the algorithm used for monitoring and correction of the timing, 
and present example results for time drifts between AERA stations as observed with the beacon method.

\subsection{Setup description}

The AERA reference beacon is an emitter of continuous waves installed about
$3\,$km west of the antenna array at the Coihueco telescope building of the Pierre Auger Observatory,
which is located at a hill overlooking AERA (see Fig.~\ref{fig:CR_aeramap}). 
The beacon system consists of a signal generator with amplifier connected 
via coaxial cable to a passive antenna emitting the signal. The signal 
generator mixes 4 sine wave signals generated by 4 Temperature Compensated Crystal Oscillators 
with a frequency stability of a few ppm. Several attenuators, filters and 
amplifiers make the overall amplitude as well as the strength of each 
individual signal adjustable, and suppress higher order artifacts due 
to non-linearities in the active electronic components. The resulting emitted signal contains the 4 desired sine waves transmitted by one single coaxial cable to the emitting antenna, 
and any other signals resulting from mixing or higher orders are suppressed by at least $40\,$dB and can be completely neglected.

The four frequencies have been chosen to lie inside the measurement 
band of 30--80~MHz of the AERA detector stations: $58.887\,$MHz, 
$61.523\,$MHz, $68.555\,$MHz, and $71.191\,$MHz. 
The choice of the four frequencies has been done such that any three of them are sufficient to 
correct for timing drifts of at least $\pm 80\,$ns, which had been observed in an earlier version of 
the AERA station electronics. In an ideal case without measurement uncertainties, two frequencies would 
suffice. Thus, the beacon system brings redundancy for signal disturbances by background.

The values of the frequencies are adapted 
to the sampling rates of the AERA stations ($180$ and $200\,$MHz) and 
the recorded trace length of at least $2048$ samples such that they 
are almost an exact match to frequency bins of the Fourier transform of 
recorded AERA time traces. By this choice, artifacts such as aliasing are 
minimized. Moreover, the beacon frequencies can be filtered both in the local 
data acquisition of each antenna station as well as in offline data analysis. Hence,
the beacon has minimal impact on self-triggering capabilities
on air-shower pulses, and does not disturb the physics analyses.

The emitting dipole antenna is horizontally aligned, almost 
parallel to the north-south aligned antennas of the AERA stations. 
Thus, the beacon signal is mainly detected in the north-south channel 
of each AERA station, and only marginally visible in the east-west channel. Since there is only one GPS clock per station, performing the timing correction on only one channel is sufficient.

In addition to its capabilities for time calibration explained in the following, the beacon, 
as a known transmitter with well-defined properties, is a powerful tool for general monitoring of the detector 
performance. As side-remark we just report two cases in which the beacon was helpful to find 
and correct mistakes. During deployment of AERA we could quickly 
identify a few stations where accidentally the two channels had been 
switched, just by comparing the beacon amplitude in both channels. 
Also, a bug in the firmware of the stations caused a slightly worse 
timing precision for events triggered externally by the surface detector of the Pierre Auger Observatory 
compared to events self-triggered on radio pulses recorded by the antenna stations. Using the beacon analysis these problems were identified 
and could be solved.

\begin{figure}[t]
  \centering
  \includegraphics[width=0.6\textwidth]{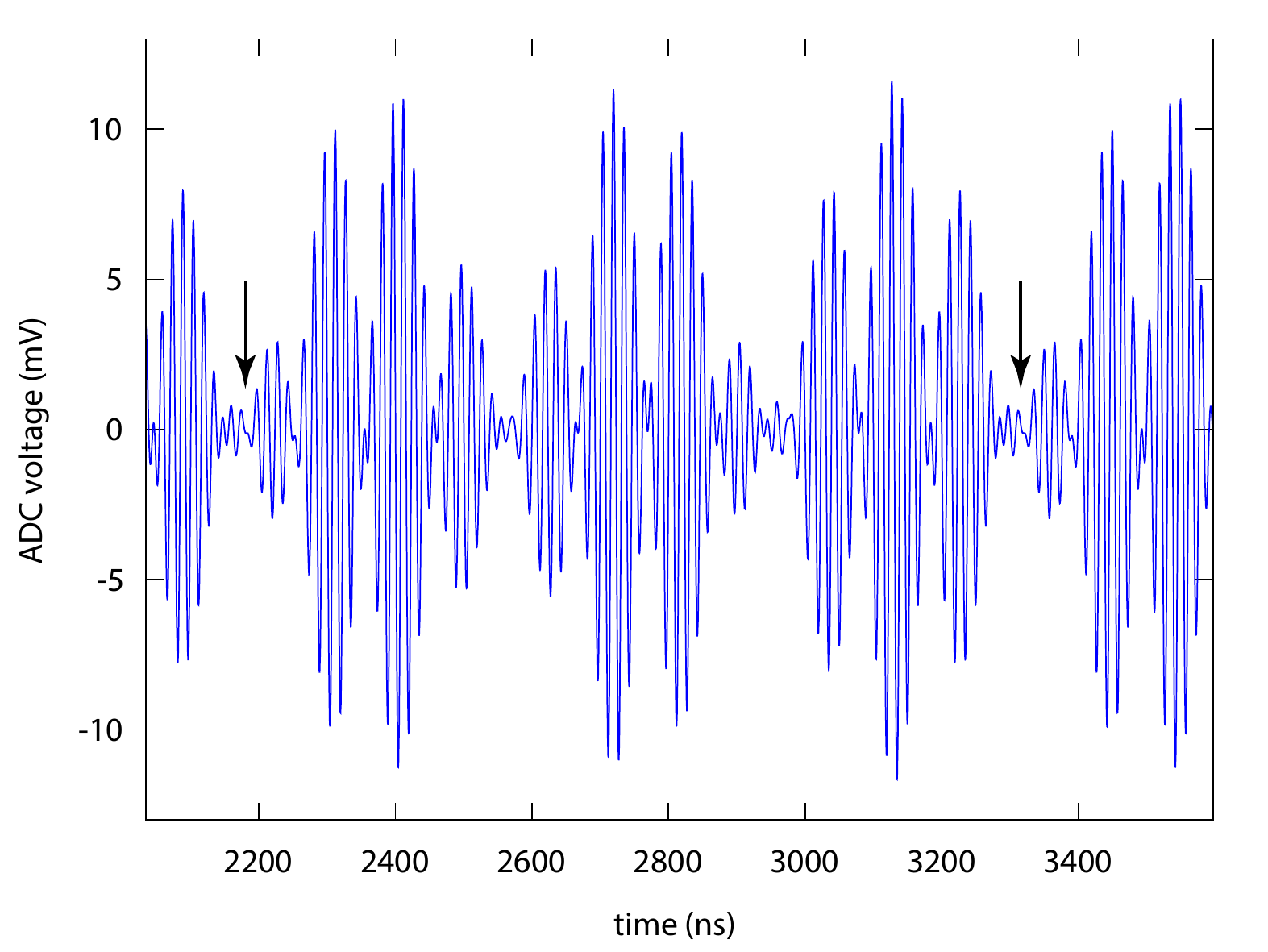}
  \caption{Beacon beat: a voltage-time trace recorded with the analog-to-digital converters (ADCs) of the north-south 
  channel of an AERA station is digitally filtered such that it 
  contains only the four beacon frequencies. The signal shape of the 
  beacon beat has a periodicity of approximately $1100\,$ns, as 
  indicated by the two arrows.}
  \label{fig:CR_beaconTune}
\end{figure}

\subsection{Method and algorithm}

The four sine waves emitted by the beacon superpose to a 
characteristic beat, which repeats approximately every $1.1\,$\textmu s 
(see Fig.~\ref{fig:CR_beaconTune}), i.e., a time series recorded by AERA of typical length $\ge 10\,$\textmu s contains several repetitions. 
During digital, offline analysis, 
the beacon beat can be isolated from other signals by applying 
digital filters to the beacon frequencies in the frequency spectrum, 
which is obtained by a Fourier transform of the time series recorded by 
the AERA antenna stations. The beacon transmitter thus does not degrade the quality of 
broad-band cosmic-ray radio pulses which AERA has been designed to 
measure.

Due to the different distances between the beacon and each of the 
antenna stations, the propagation time of the beacon signal $\tau_\mathrm{geo}$ to each 
antenna is different. Since all positions have been measured with an 
accuracy of better than $10\,$cm (slightly worse for altitude) by differential GPS, the expected 
propagation times can be calculated with an accuracy of better than 
$1\,$ns. Choosing one arbitrary station as a reference, 
it is easy to calculate the expected time difference $\Delta \tau_\mathrm{geo}$ of the beacon beat at each station with respect to this reference station (up to several \textmu s depending on the distance). 

In reality, the measured time difference of the beacon beat deviates 
from the expected time difference by typically a few times $10\,$ns (cf.~section \ref{sec_beaconResults}). The 
reason for this is the drifting behavior of the GPS clocks. However, 
since the deviation is significantly smaller than the period 
of the beacon beat, it can be unambiguously determined and corrected 
for\footnote{If fewer beacon frequencies were used, the beat period would be smaller, and thus the possible range for the correction of the 
GPS clock deviations would shrink.}. For each measured event a different, arbitrary station can be 
chosen as the reference, so that it is possible to determine the 
relative time offsets of the GPS clocks event-by-event and correct for 
them in offline analyses.

A question of practical relevance is the algorithm for the 
determination of these time differences between the beacon beat in different stations. A robust, but very time consuming way is provided by a cross-correlation analysis implemented in the offline analysis software. 
By an iterative search with sub-ns step size in a realistic range, e.g., $\pm 50\,$ns around the expectation, we determine which time shift between two stations maximizes the cross-correlation. This is the value by which the GPS time of a station has to be corrected to match that expected with the measured time of the beacon beat.

However, in practice, this procedure would 
dominate the total time needed for analyzing air-shower events. 
Therefore, we use a slightly different method still relying on the same 
principle. The method used in our offline analysis exploits the phase 
differences of the beacon sine-wave signals. 
These phases are easy to measure, since the Fourier transform applied anyway during data analysis gives not only the amplitude, but also the phase at each frequency bin. 
Mathematically, the method relying on phase differences can be derived from the cross-correlation method in the following way. 

\subsubsection{Cross-correlation of the beacon signal}

Let $s_1(t)$ be the time series measured with antenna station 1, and $s_2(t)$ the time series measured with antenna station 2. The signal of station 2 is shifted by $\Delta \tau$ to compensate for the different arrival times of the beacon beat. Then, the cross-correlation $CC$ of both signals is\footnote{In practice, the integral is calculated as a discrete sum with boundaries, which is unimportant for the theoretical treatment here.}:
\begin{align}
CC = \int s_1(t) s_2(t - \Delta \tau) \mathrm{d}t.
\end{align}

We know that the beacon signal at the two stations is the sum of the four transmitted sine waves, which have different amplitudes $A_{i,j}$  and phases $\phi_{i,j}$ at each station, but the same frequencies $\omega_j$ ($i = 1, 2$ is an index running over the stations; $j = 1, 2, 3, 4$ is an index running over the beacon frequencies):
\begin{align}
s_i(t) = \sum\limits_{j} A_{i,j} \cdot \sin (\omega_j t + \phi_{i,j}).
\end{align}

Evaluating the cross-correlation integral, all products of sines with different frequencies are 0, and only the products for sines with equal frequency remain:
\begin{align}
CC &= \smash{\sum\limits_{j} \int  A_{1,j}\ A_{2,j}\ \sin (\omega_j t + \phi_{1,j}) \cdot} \notag\\
&\qquad\qquad \cdot \sin (\omega_j (t - \Delta \tau) + \phi_{2,j})\ \mathrm{d}t.
\end{align}

As the absolute phase is not meaningful, only phase differences are 
determined experimentally and used in the calculation:
\begin{align}
\Delta \phi_j \coloneqq \phi_{2,j} - \phi_{1,j}.
\end{align}

Finally, to find the values for $\Delta \phi_j$ and $\Delta \tau$ for 
which the cross-correlation is maximized, all constant factors can be neglected in the integral:
\begin{align}
CC &= \smash{\sum\limits_{j} \int  A_{1,j}\ A_{2,j}\ \sin (\omega_j t) \cdot} \notag\\
&\qquad\qquad \cdot \sin (\omega_j (t - \Delta \tau) + \Delta \phi_j )\ \mathrm{d}t \notag\\
&\propto \sum\limits_{j} A_{1,j} A_{2,j} \cos (\Delta \phi_j - \omega_j \Delta \tau),  \label{eq_beacon}
\end{align}
which holds exactly when the integration time is an integer multiple of the periods of all beacon frequencies, or approximately when the integration time is large compared to the period of the smallest frequency.

This equation can be interpreted in physical terms. The cross-correlation is maximum when the measured phase differences between the two stations at all beacon frequencies correspond exactly to what is expected from the time difference of the beacon beat at the two stations. Moreover, the different beacon frequencies receive a weight according to the amplitudes measured by the stations.

Since the amplitudes $A_{i,j}$ and the phase differences $\Delta 
\phi_j$ are measured quantities, the only variable in equation 
\ref{eq_beacon} is the time shift $\Delta \tau$. A starting guess for $\Delta \tau$ is
$\Delta \tau_\mathrm{geo}$, which is the geometrically expected difference in the propagation time of the beacon signal. Afterwards, we 
search for the $\Delta \tau_\mathrm{max}$ providing the global minimum 
of equation \ref{eq_beacon} in a range of $\pm 100\,$ns around the expectation $\Delta \tau_\mathrm{geo}$. The difference between $\Delta \tau_\mathrm{max}$ and $\Delta \tau_\mathrm{geo}$ is interpreted as current offset of the GPS clocks relative to each other, and used as correction value.

\subsubsection{Practical implementation}

In practice two major problems occur, which are not reflected by the pure mathematical derivation, but are taken into account for the implementation of the beacon method.

First, the phase measurements are affected by noise. Since the orientation of the electric field vector 
is random for noise, thus, it generally points in a direction different from that of the electric field vector of the signal.
Consequently, the random phase of noise changes the measured phase of the signal. 
For ns-timing precision, a maximum phase error of $\lesssim 20^\circ$ can be tolerated. 
The $20^\circ$ correspond to a shift of slightly less than $1\,$ns, 
since a full period ($\widehat{=}~360^\circ$) is about $17\,$ns for the lowest beacon frequency. 
On the basis of simple trigonometry it is possible to estimate the minimum signal-to-noise ratio, which is tolerable 
in order to avoid phase errors of more than $20^\circ$: 
the phase error is maximum, when the electric field vector of the noise is perpendicular to the electric field vector of the beacon signal. 
Then the phase error is $\arctan (N/S)$ with $S$ and $N$ the lengths of the signal and noise vectors, respectively, which requires 
the signal amplitude to be $\tan 20^\circ = 2.75$ times larger than the noise amplitude.
As usual in radio engineering, the signal-to-noise ratio is defined in the power domain (amplitudes squared), 
such that the threshold for the signal-to-noise ratio is $2.75^2 \approx 8$.
In the case of the AERA beacon, the emission power has been set to exceed this threshold. 
In fact, with the exception of a few measurements with exceptionally strong background fluctuations, 
the signal-to-noise ratio is much larger ($>100$ for stations close to the beacon). 
Then, systematic uncertainties dominate, which have not been 
investigated in detail, since they appear to be unimportant for the 
desired ns-timing precision. Thus, weighting the different 
beacon frequencies by their amplitude as in equation \ref{eq_beacon} is unreasonable. Consequently, the 
weighting by the amplitudes is replaced by a weighting with the 
signal-to-noise ratio, capping the weight to a maximum value for any signal-to-noise ratio $\ge 10$.

Second, even for very large signal-to-noise ratios the phase is not 
completely determined by the geometrical propagation of the beacon 
signal. Ground properties as well as the antenna characteristics 
influence the measured phases. These effects are very hard to calculate 
or simulate, since the beacon signal propagates nearly horizontally, 
parallel to the ground. Thus, the ground is within the first Fresnel 
zone of the signal and cannot be neglected. Therefore, usual free-space approximations in the antenna simulations do not hold. 
Moreover, measurements of the group delay are cumbersome, and have been done only for selected arrival directions to cross-check the
antenna simulations \cite{AERAAntennaPaper2012}. Generally, the used antenna model compares well to the measurement, 
but with larger deviations for near-horizontal signals.
In particular, the simulated group delays of the antennas 
deviate significantly from measurements for the zenith angle of the beacon signals, 
which is $\theta \approx 89^\circ$ decreasing slightly with increasing distance of the antennas to the beacon. 

Hence, in the actual timing correction performed event-by-event 
during analysis, the measured phases are not compared to the calculated 
expectation $\omega_j \Delta \tau_\mathrm{geo}$, but instead to 
reference values derived from phases measured by the real antenna 
stations. These reference values are obtained by averaging over 
approximately $100$ triggered AERA events corresponding to the statistics of several hours. 
The measured reference values are in agreement with the geometrically calculated values. 
However, for some stations and frequencies the deviation is slightly larger than the desired accuracy of 
$1\,$ns. Consequently, it is indeed necessary to use the measured 
reference values instead of the calculated ones. 

\subsection{Results of beacon analysis}
\label{sec_beaconResults}

In the standard analysis pipeline of AERA, the beacon correction is 
automatically applied for all recorded events. This means that the 
timing of AERA stations participating in the event is shifted by the 
correction value obtained from the beacon analysis. 
A statistical study for many events of the amount by which one station has to be shifted against a specific other station reveals the relative timing accuracy and stability. 
To enhance the statistics, the beacon timing correction is not only applied to cosmic-ray candidate events, 
but also on background data triggered periodically at least every $100\,$s, 
and we checked that the timing of cosmic-ray events is not systematically different from that of background measurements. 

\begin{figure}[t]
  \centering
  \includegraphics[width=0.6\textwidth]{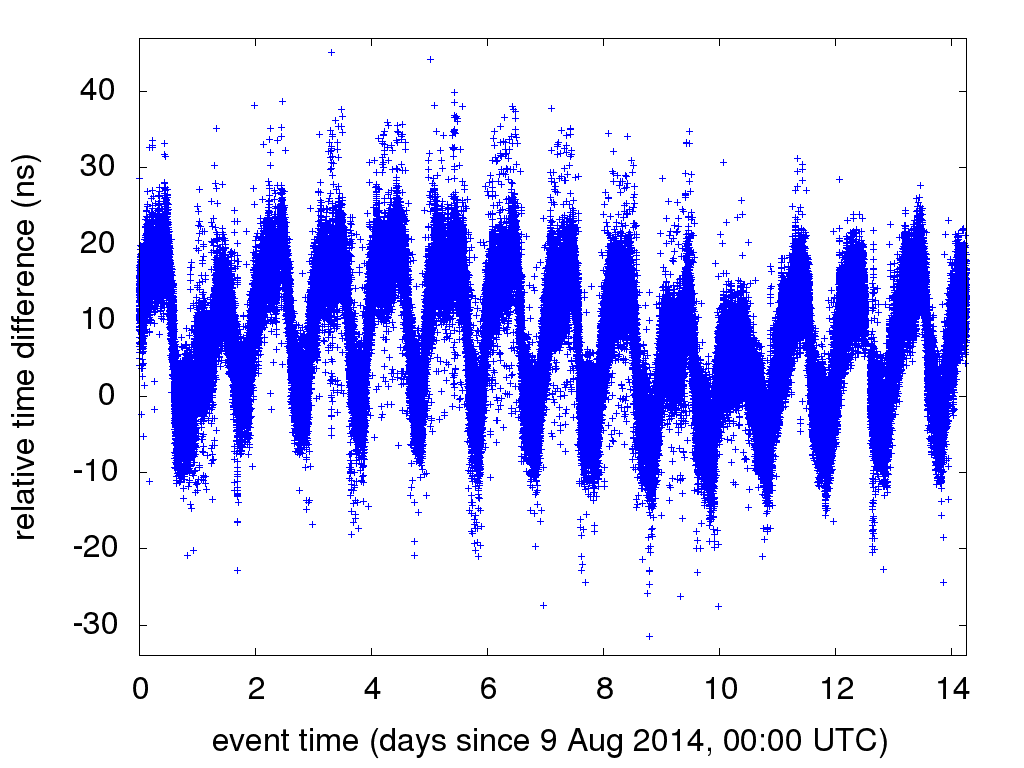}
  \caption{Example of the relative timing between two AERA stations 
  determined with the beacon method. Each entry denotes the size of 
  the timing correction between two typical AERA stations in one 
  measured event, interpreted as a time offset between the GPS clocks 
  of the two detector stations. The few outliers have not been investigated 
  in detail, since the focus of the present analysis is on the proof-of-principle 
  and the general behavior of the relative timing. A possible explanation for the outliers might be  
  mis-measurements, e.g., because of occasional high background disturbances.}
  \label{fig:GPSdrifts}
\end{figure}

Fig.\ \ref{fig:GPSdrifts} shows a graphical representation of such a 
statistical study of the relative timing between two stations over 
several days. At a random point in time, the time offset between two 
AERA stations can be as large as several times $10\,$ns. This would dominate 
the total timing accuracy and spoil any interferometry or wavefront 
analysis, if not corrected for. Moreover, the time offset between two 
stations experiences systematic drifts on time scales of several hours. 
The relative timing is stable only over short periods of a few minutes  
at a level of a few ns, corresponding to the nominal precision of the GPS 
clocks. However, even over these short time 
periods two GPS clocks of two different detector stations can still
have an arbitrary offset of a few times $10\,$ns.

The reason for these drifts and offsets is not known. 
The day-night structure in Fig.\ \ref{fig:GPSdrifts} indicates that environmental effects, like temperature, play a role,
but they cannot fully explain the drifts. We performed 
several cross-checks to exclude possible explanations. In particular, the 
effect cannot be explained by the accuracy of the location stored 
in the GPS clocks. In laboratory measurements we connected two equal 
clocks to the very same antenna, and still observed qualitatively the 
same drifts, though on a slightly smaller scale, but still exceeding the $5\,$ns accuracy claimed by the manufacturer. 
This also excludes any changes of environmental conditions as the sole reason for the drifts, since 
both clocks have been exposed to exactly the same conditions in the 
laboratory. Thus, environmental effects in the field apparently do not cause the drifts, but enhance 
their size, which would explain the rough day-night structure in the offsets. 
The size of the drifts in the AERA stations was successfully reduced, by the 
installation of GPS antennas with higher gain, indicating that the 
signal-to-noise ratio of the received GPS signals plays a role (the results in Fig.\ \ref{fig:GPSdrifts} are with the new GPS antennas). 
Also the electronics of the local data acquisition seem to influence the size of 
the drifts, possibly by introducing some local interference. It seems 
that a complex combination of effects causes the 
observed drifts.

Fortunately, the GPS clock drifts can be 
corrected for in the offline analysis of measured events, regardless of their physical origin. However, we have to independently 
verify whether the drifts determined and corrected for by the beacon are indeed 
intrinsic to the GPS clocks and not caused by some external effect such as 
environmental influence on the propagation delay of the beacon 
signals. To verify the beacon correction, and to cross-check its 
finally achieved accuracy, we thus developed an independent time 
calibration using airplane signals described in the next section.


\section{Airplane calibration} \label{sec:airplane}

In this section, we describe the general concept, the hardware and software 
setup, and the analysis strategy for the timing calibration using 
commercial airplane signals.

\subsection{Method description}

To use a source for timing calibration, two requirements need to be 
fulfilled. First, the source must emit a signal that can be 
identified in the data taken by the radio detector array, 
simultaneously in as many detector stations as possible. A short, 
bandwidth-limited pulse is ideal as it can be time-tagged precisely.
Moreover, the position of the source needs to be known at the time of emission. 
From the known position and the relative arrival time of the signals in 
different detector stations, the time offsets between different 
detector stations can then be determined provided that the model for the wavefront used for the reconstruction is well adapted to the actual wavefront.

These requirements are fulfilled for commercial airplanes. Surprisingly, 
(some) commercial airplanes emit radio pulses in the 30-80~MHz frequency range 
covered by AERA. The origin of these radio 
pulses is unclear.
In 2011, using an upgraded version of a small prototype radio 
array at the Pierre Auger Observatory~\cite{RAuger}, we regularly 
observed airplane signals in the frequency band 30-260~MHz. 
However, the characteristics of these signals varied from one airplane to another. In 
some cases, the signals were periodic with a very high level of 
accuracy (repetitive bunches of 4 transients separated by $132870\pm 
90~\mu$s followed by a pause of $861120\pm 280~\mu$s). In other cases, 
the arrival times of the signals did not exhibit any clear structure. 
Similar observations were made in CODALEMA.
Possible sources of emission could be the TCAS (Traffic Collision 
Avoidance System, periodic emission), the ADS-B~\cite{ADS-B} 
transponder (periodic emission), the airplane navigation lights 
(periodic) or the DME (Distance Measuring Equipment, aperiodic) 
\cite{privcomm}. No definitive answer could be found as the next step 
for determining the source would require classified information that we 
could not obtain. It is, however, not necessary to know the 
mechanism for the emission or the exact characteristics of the 
radio pulses to use them for timing calibration.

The position of an airplane can be readily determined in real time from digital broadcasts 
that modern airplanes emit at rates of $0.5-1\,$Hz, via the ADS-B service.
These digital broadcasts at a frequency of 1090~MHz contain 
information such as an airplane-specific call-sign, longitude and latitude, altitude, 
velocity and heading of the airplane. They are intended to inform 
other aircraft and air traffic control in real time of airplane 
positions, but can indeed be received easily with equipment 
costing less than 20 USD by any interested person. All that is needed 
is reception hardware for the frequency of 1090~MHz which can be used
as ``software-defined radio'',  meaning that demodulation, decoding and interpretation of the 
received signals are performed in software on a computer. Inexpensive receivers with a USB 
interface originally intended for the reception of digital TV signals 
(DVB-T) suit this purpose. 
There is a broad community employing such USB receivers with RealTek RTL2832U chipsets for the reception of various 
analog and digital signals under the term ``RTL-SDR''.

\begin{figure}[t]
  \centering
  \includegraphics[width=0.5\textwidth]{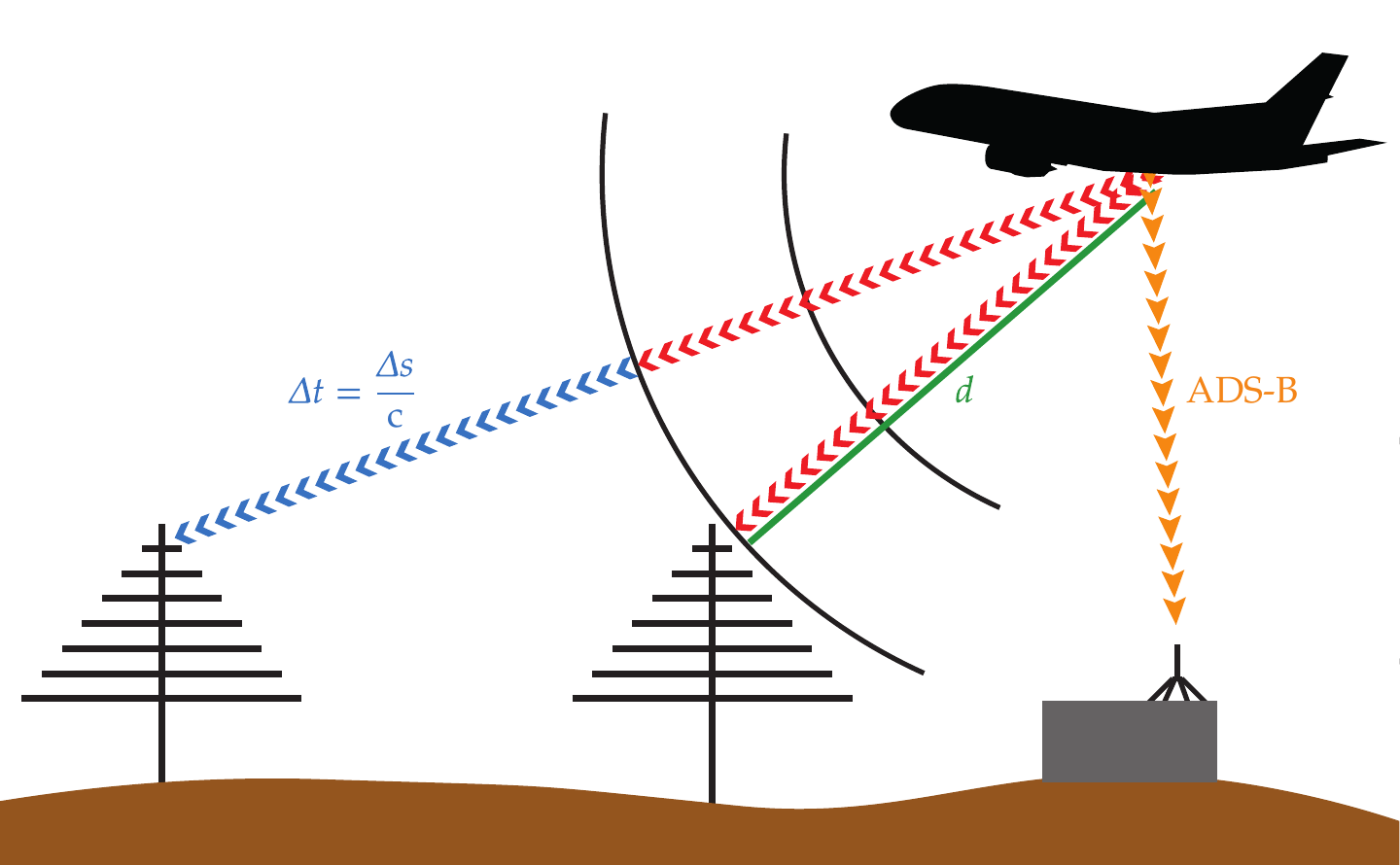}
  \caption{Concept of the time calibration using commercial airplanes. The
    airplane broadcasts its position via digital ADS-B packets at 
    1090~MHz. These signals are received and interpreted by a 
    dedicated setup in the AERA field in real time. In addition, 
    (some) commercial airplanes emit pulsed signals in the frequency 
    range of 30--80~MHz recorded by the AERA detector stations.}
  \label{fig:airplanes_skizze}
\end{figure}

By combining the real-time position information from ADS-B data and the 
radio pulses emitted by the airplane as recorded in the AERA 
data stream, time offsets between GPS clocks of individual AERA 
stations can be calculated.
The overall concept is visualized in Fig.\ \ref{fig:airplanes_skizze}. Using 
the known position of both the airplane (interpolated from the ADS-B messages to the exact time of the event)
and the AERA detector stations, 
the expected arrival times of the radio pulses 
can be calculated. The typical values are on the order of several $10\,$\textmu s, since 
the typical distances are on the order $10\,$km (explanation in more detail follows later). 
Deviations from these expected arrival times can 
be interpreted as timing drifts between pairs of detector stations.

An additional complication that needs to be addressed is that the AERA central data acquisition system, 
which triggers on radio pulses detected by AERA stations as well as information from external 
detectors, has been optimized to suppress anthropic radio sources. In 
particular, it blinds regions of the sky from which radio pulses have 
been received repeatedly within a few minutes.  Airplanes crossing the 
sky over the course of a few minutes constitute such 
sources and would thus be filtered out by our data acquisition. 
Therefore, the data acquisition software had to be modified, as will 
be described below.

\subsection{The setup}

A simple hardware setup was deployed at the Central Radio Station of 
AERA (cf.\ Fig.\ \ref{fig:CR_aeramap}) to record the ADS-B information transmitted by airplanes in 
real time. It consists of a tailor-made antenna for the 1090~MHz ADS-B 
signals, a USB receiver connected to a laptop and dedicated software 
to use the receiver as a software-defined radio decoding and processing 
the ADS-B data packets. We will shortly describe the individual items in 
the following.

\begin{figure*}[t]
      \centering
      \includegraphics[height=7.5cm]{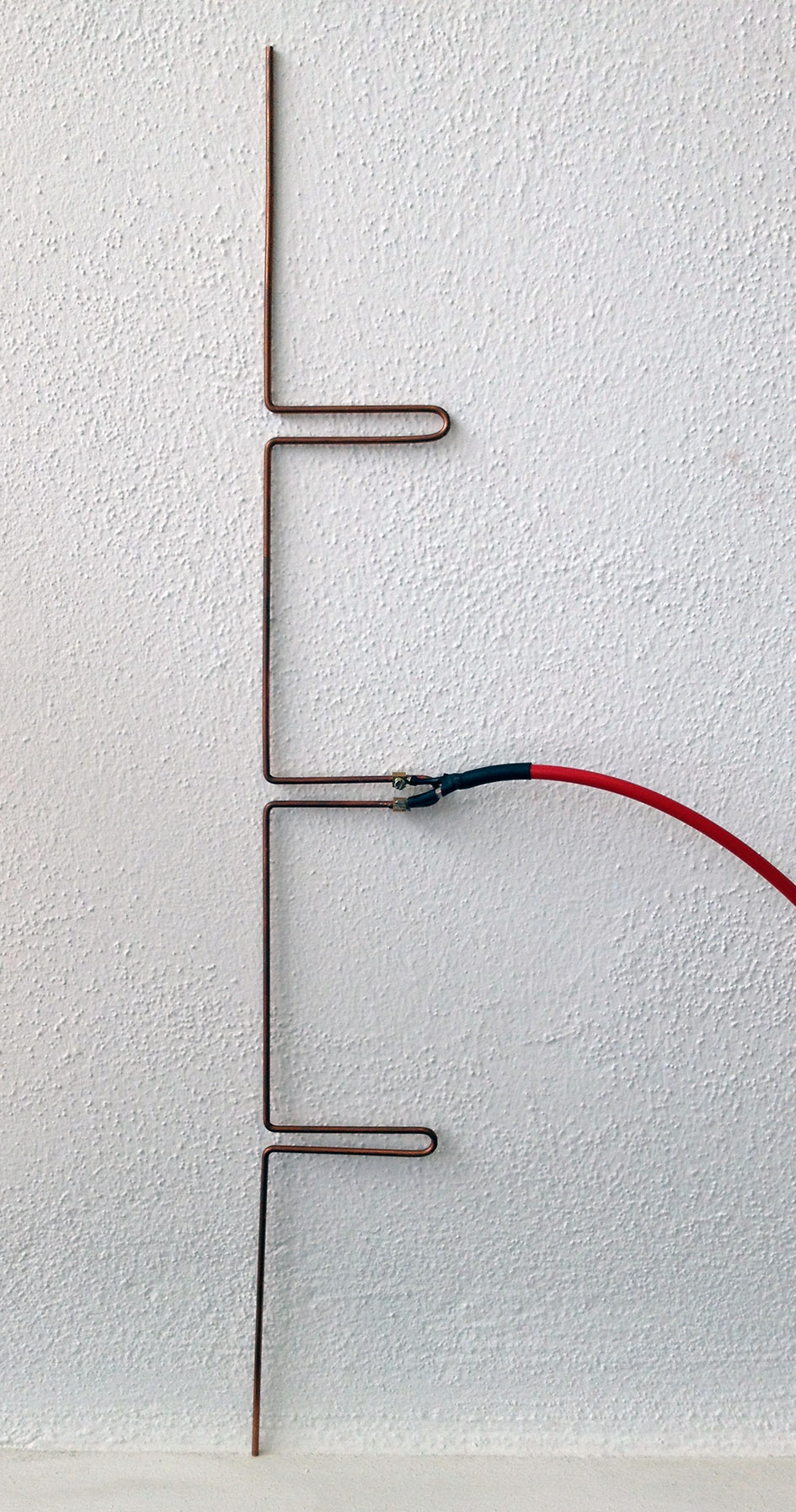}
      \includegraphics[height=7.5cm]{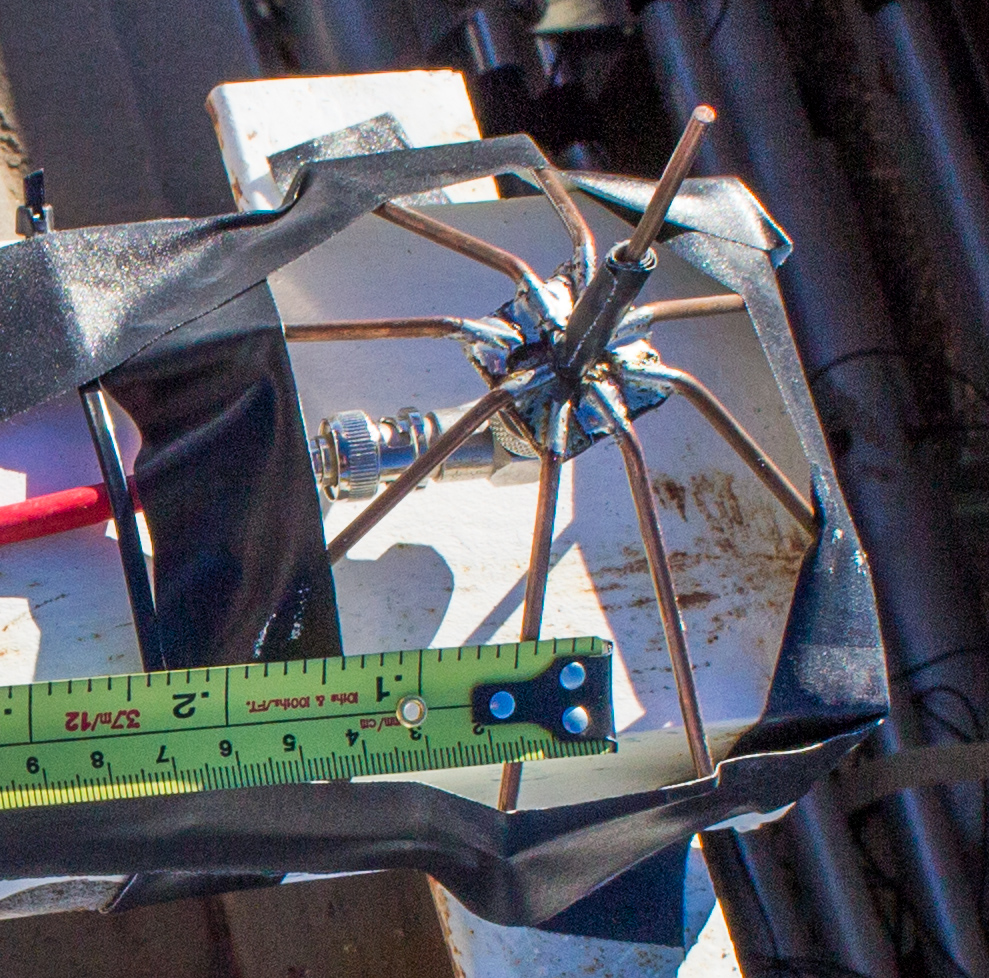}
      \caption{ADS-B antennas of two different types: four-segment 
      collinear antenna (left) and ground-plane antenna (right). 
      The ground-plane antenna was deployed in the AERA field.}
      \label{fig:airplanes_antennas}
\end{figure*}

\begin{figure*}
      \centering
      \includegraphics[width=0.45\textwidth]{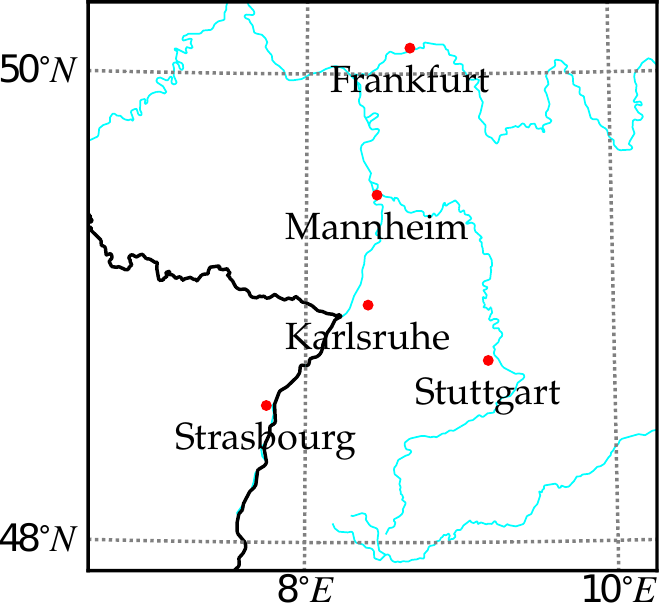}
      \hfill
      \includegraphics[width=0.45\textwidth]{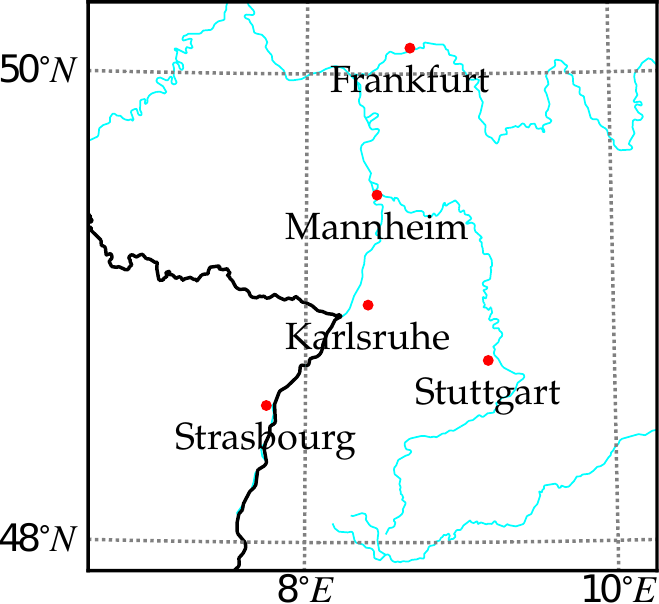}
      \caption{Reception area (in green) of the two ADS-B
          antenna tested at Karlsruhe Institute of 
          Technology: Four-segment 
          collinear antenna (left) and ground-plane antenna (right). 
          Each single green point denotes a received airplane 
          position. The western hemisphere was shielded by a concrete 
          building.}
        \label{fig:airplanes_antennas_range}
\end{figure*}

After initial tests with a simple monopole antenna, two antennas optimized for 1090~MHz signals were 
built and tested. The two types, 
a four-segment collinear antenna and a ground-plane antenna 
are depicted in Fig. \ref{fig:airplanes_antennas}. The latter antenna 
was built according to a design published by hobby 
enthusiasts\footnote{http://www.diylightanimation.com/wiki/images/0/01/\\DIY\_Quarter\_Wave\_Antenna.pdf}. 
Tests conducted at the Karlsruhe Institute of Technology (Fig.\ 
\ref{fig:airplanes_antennas_range}) showed that the 
ground-plane antenna reproducibly sees airplanes to further 
distances than the collinear antenna (or a simple monopole, not shown 
here). It was thus chosen as the antenna that was deployed on the roof of the Central Radio Station
within the AERA field. In spite of its improvised design, the reception quality of the ground-plane 
antenna fully satisfied our needs; we only need to reliably detect 
airplanes in close vicinity to AERA. Further optimizations were thus
not carried out.

The ground-plane antenna installed in the AERA field was connected to a USB 
receiver with a RealTek RTL2832U chipset and a Rafael Micro R820T tuner chip which can be 
tuned to 1090~MHz. This USB receiver, connected to a commercial laptop, has been 
working reliably over months of continuous operation.

For the decoding and interpretation of the ADS-B data packets, the 
open-source software package ``dump1090''\footnote{https://github.com/MalcolmRobb/dump1090} was extended for our needs. In 
addition to decoding the position and velocity information of airplanes 
for which ADS-B packets have been received, the software identifies airplanes that come close 
enough to AERA to be visible at a zenith angle up to 80$^{\circ}$ from the vertical.
While an airplane fulfills this condition, notification packets 
detailing the relevant information are sent in real time to our data acquisition 
software which then accepts radio triggers from a window of 
5$^{\circ}$ around the known airplane position instead of suppressing 
these radio pulses of clearly anthropic origin. In parallel, a log file with decoded ADS-B information 
such as latitude, longitude and altitude is stored on disk for easy 
follow-up analyses. The ADS-B messages do not contain reliable information on the position uncertainty, 
but our analysis itself confirms that the information is accurate enough to apply the presented method 
(cf.~the cross-check of the beacon and airplane methods in section \ref{sec:combined}).

\subsection{Airplane calibration analysis}

Here, we explain how the analysis of airplane signals is 
carried out to deduce the time offsets of GPS clocks in individual 
radio detector stations with respect to one reference station.

When the central data acquisition is notified of an approaching 
airplane, it will let radio pulses from corresponding arrival 
directions trigger the readout of all detector stations, marking the 
corresponding events with a special ``airplane trigger'' flag. This means that 
airplane signals are detected by the same self-trigger algorithm also used for air showers,
and can be separated from the normal AERA data stream using this flag .

\begin{figure}
  \centering
  \includegraphics[width=0.5\textwidth]{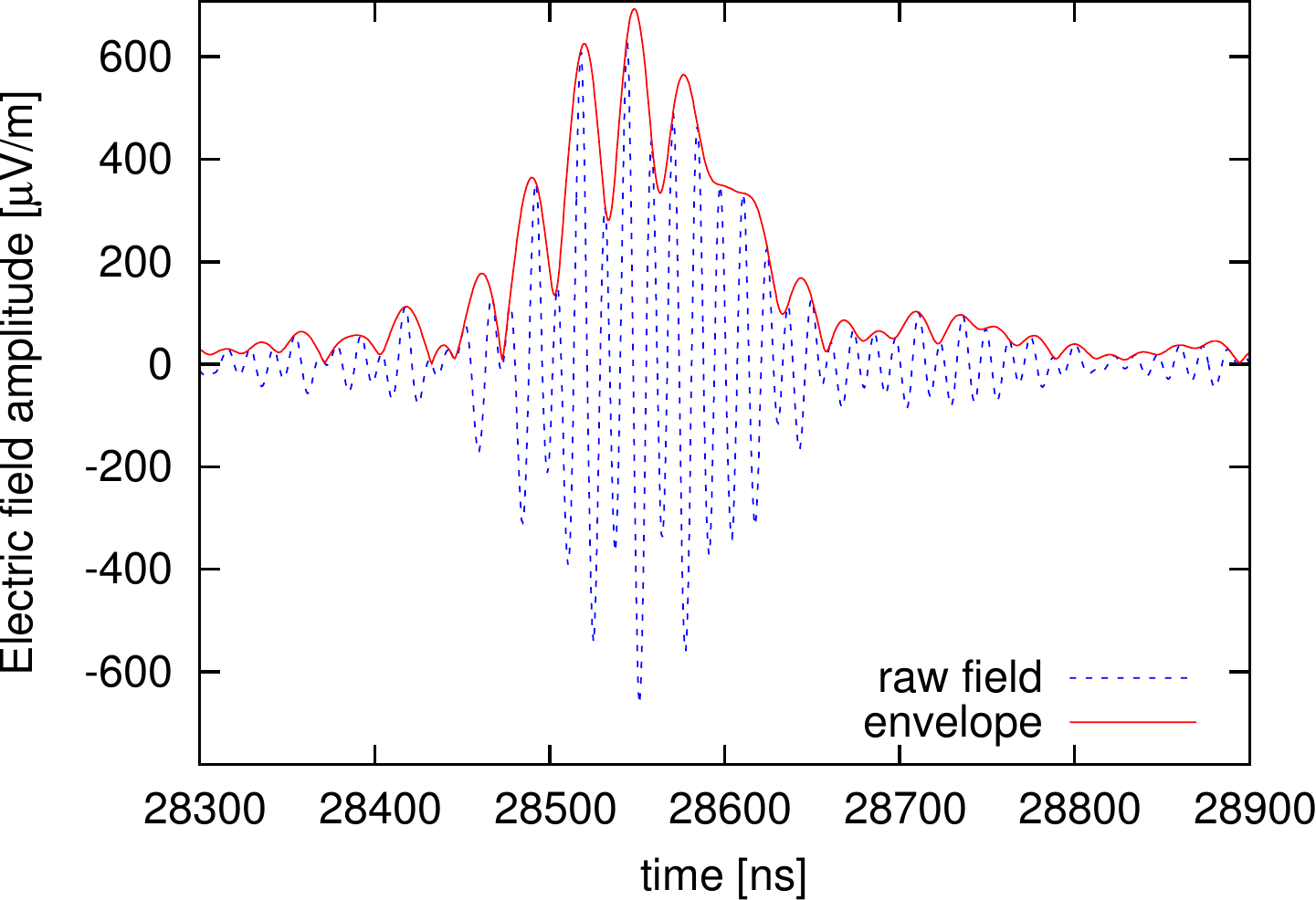}
  \caption{Typical radio pulse received from an airplane in the 
  north-south component of the reconstructed electric field. A Hilbert 
  envelope is calculated to read off the time of the pulse maximum. 
  Time zero corresponds to the start of the recorded time trace, which has a total length of 50~microseconds.}
  \label{fig:analysis_pulses}
\end{figure}

We apply an iterative direction reconstruction 
\cite{AbreuAgliettaAhn2011} to 
these AERA events. We require simultaneous detection in at least 10 radio detector stations 
with a signal-to-noise ratio (defined as peak power divided by RMS 
noise power) of at least 15. In addition to the arrival direction, we 
reconstruct a source distance using a spherical wavefront as adequate 
for a point-like source and exclude events\footnote{The wavefront 
reconstruction is applied to AERA raw data without any prior correction 
of detector station time offsets and is thus affected by GPS clock 
drifts. However, we only use the reconstructed distance to cut away 
rare radio-frequency interference pulses clearly not originating from airplanes, not for the 
determination of time offsets. The good agreement of the direction reconstructed from AERA events and ADS-B data illustrates that the resulting sample of ``airplane events'' is very clean.} with a wavefront radius of 
curvature of less than 7.5~km (commercial airplanes travel at 
altitudes of 10~km or more) or more than 99~km. A typical airplane pulse as seen in one 
radio detector station is shown in Fig.\ 
\ref{fig:analysis_pulses}. We apply an upsampling procedure and a Hilbert envelope to the 
electric field trace reconstructed from the voltage traces of the 
east-west and north-south detection channels, respectively \cite{AbreuAgliettaAhn2011}, and use the time of the maximum of 
the enveloped signal as the pulse arrival time in the reconstruction.

\begin{figure}
  \centering
  \includegraphics[width=0.49\textwidth]{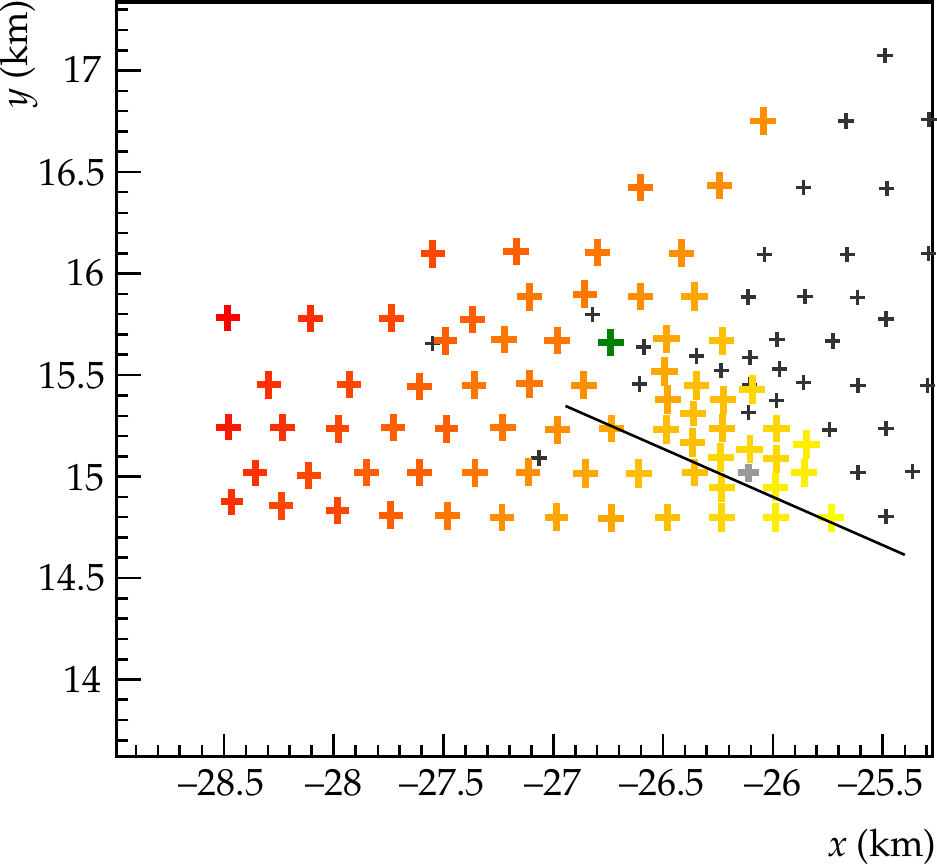}
  \caption{Typical airplane event as shown in the Offline
    EventBrowser. Every station is indicated by a cross. One station
    without detectable signal is marked with a gray cross, stations that are not in data acquisition are shown in
    black. Stations with a signal have a color indicating the arrival 
    time from early (yellow) to late (red). The reconstructed arrival 
    direction is marked with a black
    line. The station used as reference station in the analysis
    (station 40) is marked in green.}
  \label{fig:analysis_EventBrowser}
\end{figure}

\begin{figure}
  \centering
  \includegraphics[width=0.7\textwidth]{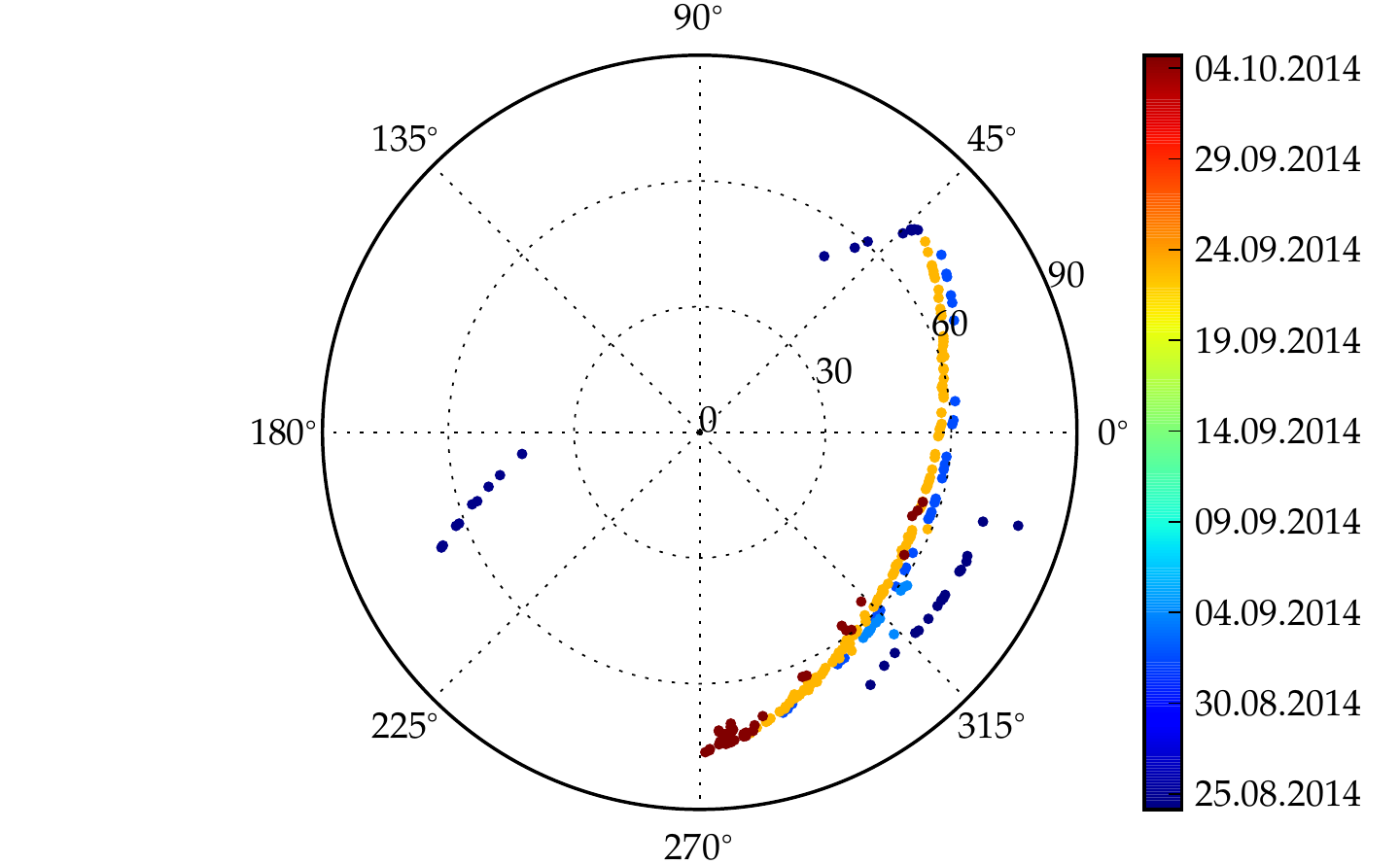}
  \caption{Reconstructed arrival directions of AERA events 
    for airplane fly-overs with at least 10 successful detections. The 
    sky map has the zenith at the center (cf.\ table 
    \ref{tab:analysis_airplaneevents} in the appendix) and East at 0$^{\circ}$. The
    color of the dots represents the date of detection. For some airplanes
    a clear trajectory is visible.}
  \label{fig:analysis_2014_ePLAN-0925}
\end{figure}

An overview of a typical airplane event is shown in Fig.\ \ref{fig:analysis_EventBrowser},
illustrating that the radio 
pulses from airplanes trigger almost all radio detector stations 
simultaneously. (The radio stations in the eastern part were being 
read out with a separate data acquisition system not supporting the airplane trigger.) The 
color code denotes the relative arrival time of the radio pulses 
at the individual detector stations; in this case it is obvious that 
the source must be south-east of the array. In Fig.\ \ref{fig:analysis_2014_ePLAN-0925} we illustrate the 
reconstructed arrival directions of all airplanes which have been seen in at least 10 AERA 
events within our data set from June to October 2014.

For airplane events reconstructible using AERA data, the corresponding recorded ADS-B data 
are analyzed. A unique airplane can be identified 
as the source of each detected AERA event. A total of 21 individual 
flights passing over AERA have been identified. Details are listed in 
table \ref{tab:analysis_airplaneevents} in the appendix. Some airplanes, uniquely 
defined by their identification code, the so-called Mode-S call-sign, 
have been spotted several times in different fly-overs.

\begin{figure}
  \centering
  \includegraphics[width=0.6\textwidth]{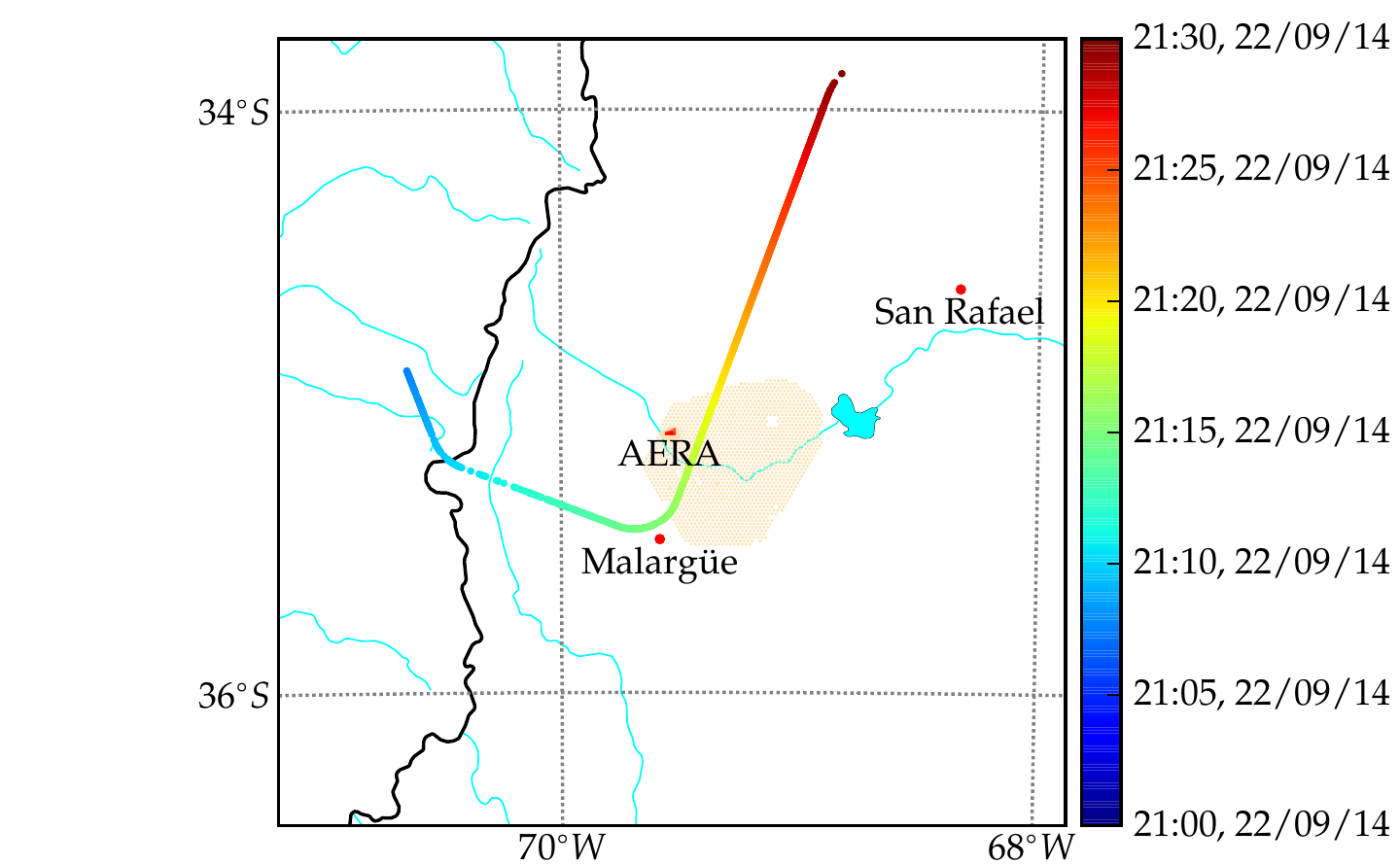}
  \caption{Trajectory of the airplane passing the AERA field (marked in
    red) and the main surface detector array of the Pierre Auger Observatory (orange shaded area) on September 22, 2014 at
    around 21:15~UTC based on ADS-B data.}
  \label{fig:analysis_arm20140922sd}
\end{figure}

\begin{figure*}
  \centering
  \includegraphics[width=0.97\textwidth]{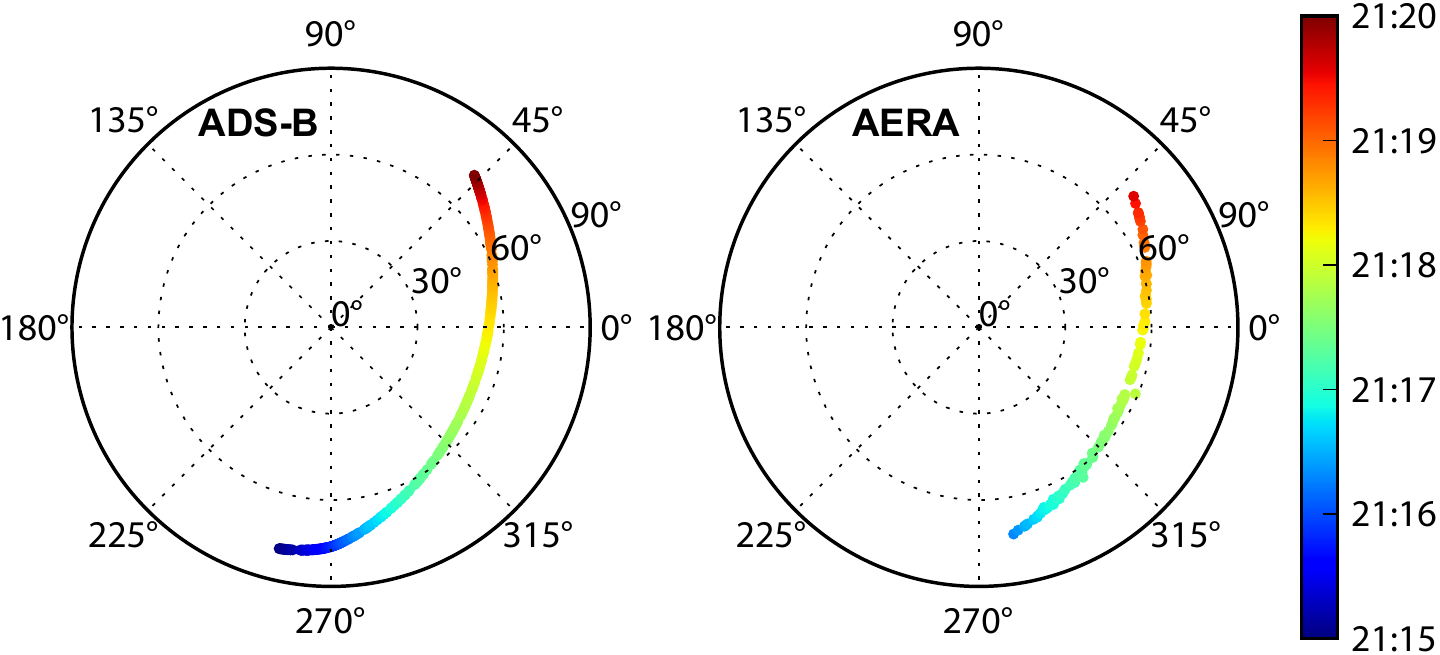}
  \caption{Position of the airplane passing nearby AERA on September 22 based on
    ADS-B data (left) and reconstructed arrival
    directions from 115 AERA events of the same airplane (right). The 
    mean deviation between the direction reconstructed with AERA and taken from the ADS-B data is $0.3^\circ$. The color code 
    denotes the UTC time at which the airplane event was detected.}
  \label{fig:analysis_20140922+adsb}
\end{figure*}

The ADS-B data for each airplane provide position 
information also for those parts of the airplane trajectory for which no AERA 
triggers were generated (be it because there were no radio pulses or 
because they came from directions with zenith angles above 
80$^{\circ}$, for which the data acquisition was not accepting 
airplane triggers). An example airplane trajectory deduced from 
ADS-B data is shown in Fig.\ \ref{fig:analysis_arm20140922sd}.  There is good agreement between 
the arrival direction reconstructed from the AERA data (30-80~MHz) and 
the position information digitally broadcast by the airplane via 
ADS-B packets (1090~MHz), as shown for the same airplane in 
Fig.\ \ref{fig:analysis_20140922+adsb}. The directions 
reconstructed from AERA data and calculated from ADS-B position data match
generally within $0.3^\circ$. Also, the source distance 
derived from the spherical wavefront fit (not shown here) is in good agreement 
with the distance calculated from the ADS-B position data. (The source 
distance reconstructed from AERA data is only used as a cross-check 
here. For further analysis, only the ADS-B position is used.)

The ADS-B position data is only available at discrete times, typically 
with a rate of $0.5-1\,$Hz. Therefore, the ADS-B position data are 
linearly interpolated in latitude and longitude. 
Surprisingly, this yields better consistency with the positions reconstructed from AERA 
data than an interpolation also taking into account the velocity and 
heading information available from the ADS-B data. The ADS-B position 
data is not only used for the calculation of the expected pulse arrival 
times, but is also used to determine an individual direction per AERA 
detector station for the application of the antenna pattern.
(The airplane is at a typical distance of 10 to 30~km, i.e., it 
is seen at angles differing on the order of $5-10^\circ$ for the different station of the array.)


As only relative time differences between AERA detector stations can be 
determined, one detector station is chosen as the reference (which by 
definition has a time offset of zero). The reference station is chosen such that 
it is near the geometric center of the radio detector array and that 
it has a positive pulse detection in a large number of the AERA events 
for a given airplane.

The light travel time from the airplane to each antenna is calculated assuming that 
the refractive index of the atmosphere has a constant value of $n \approx 1.00024$ as 
that corresponding to the altitude of AERA (1,560~m above sea level). 
The calculated arrival times are then compared 
with the actual measured pulse arrival times with respect to the 
chosen reference station. The difference between the measured and 
predicted arrival times is the time offset of the individual detector 
station with respect to the reference station.

\begin{figure}[t]
  \centering
  \includegraphics[width=0.7\textwidth]{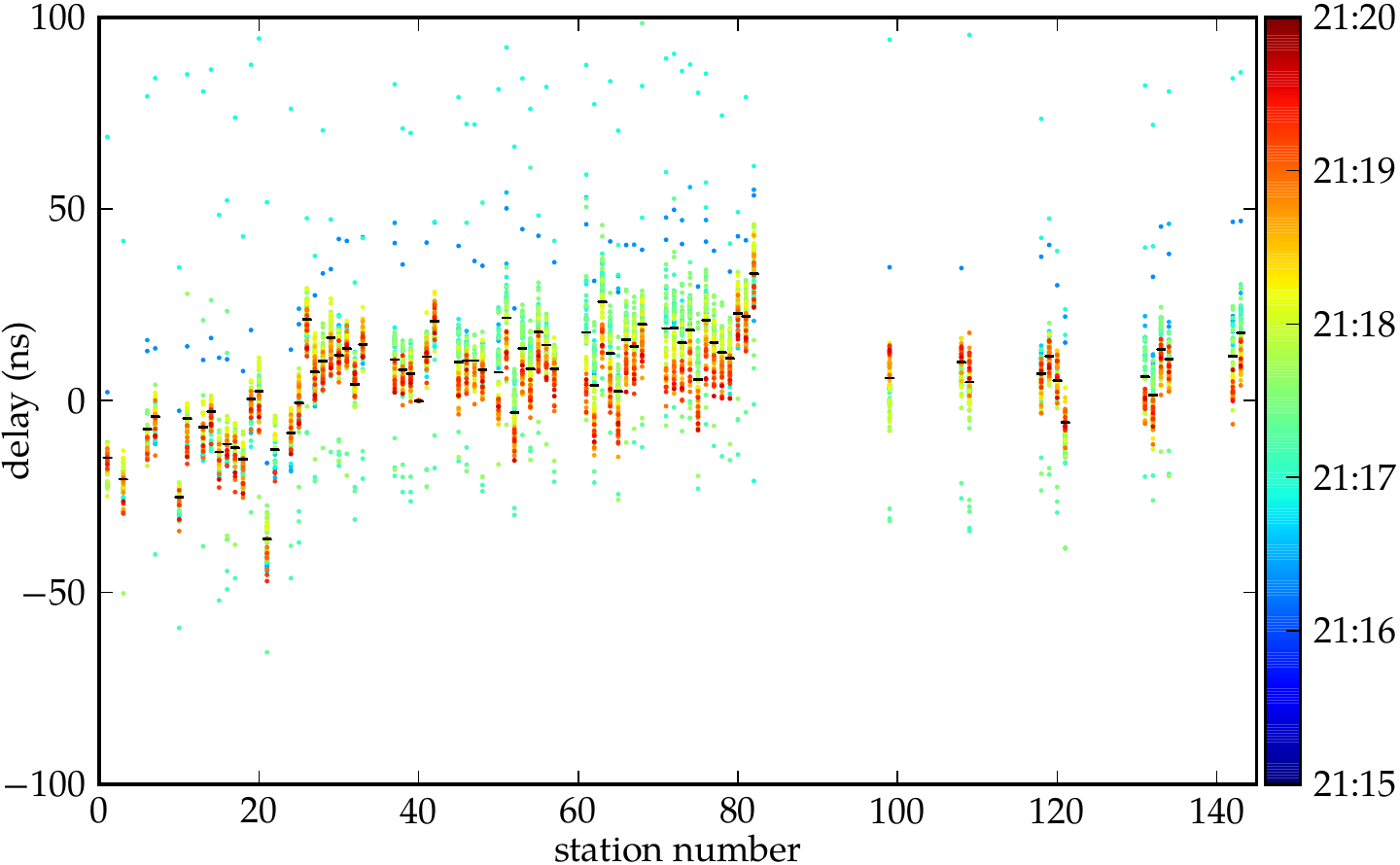}
  \caption{Per-event time offsets for all AERA detector stations 
  determined from AERA and ADS-B data for an airplane detected on 
  September 22. Time offsets are relative to reference station 40. 
  Stations 1-24 are equipped with logarithmic-periodic dipole 
  antennas, stations with numbers 25 and higher are equipped with 
  butterfly antennas. Gaps in the station numbering are related to 
  stations not participating in the airplane trigger (marked gray in 
  Fig.\ \ref{fig:CR_aeramap}). The color code denotes the local time at which the airplane event was detected.}
  \label{fig:analysis_20140922_woB}
\end{figure}

The result of this analysis for the example airplane already presented 
in Figs.\ \ref{fig:analysis_arm20140922sd} and \ref{fig:analysis_20140922+adsb} is depicted in Fig.\ 
\ref{fig:analysis_20140922_woB}. A number of features can be 
identified in the resulting time offsets. The individual offset values 
determined from the different AERA events in which the airplane was 
detected have a scatter of typically less than 10~ns, but there are 
significant outliers. Within the scatter of $\approx 10$~ns, there is a time-ordering, 
i.e., there must be some systematic effect. We will discuss both 
aspects at a later point. The mean values calculated from 
the distributions are typically within 20~ns of the zero-line (which 
would correspond to perfect time synchronization of the array). This 
is approximately the scale of the drifts previously observed by the 
beacon analysis.

\section{Combined timing analysis} \label{sec:combined}

With both the beacon timing calibration and airplane timing analysis 
in place, the two methods can now be combined to check them for 
consistency. If the beacon correction can indeed detect and mitigate 
timing drifts in the GPS clocks of the different AERA stations, then 
a follow-up airplane analysis should yield vanishing time offsets for all 
stations to the reference station.

\subsection{One airplane}
\label{sec:one_airplane}

\begin{figure}[t]
  \centering
  \includegraphics[width=0.7\textwidth]{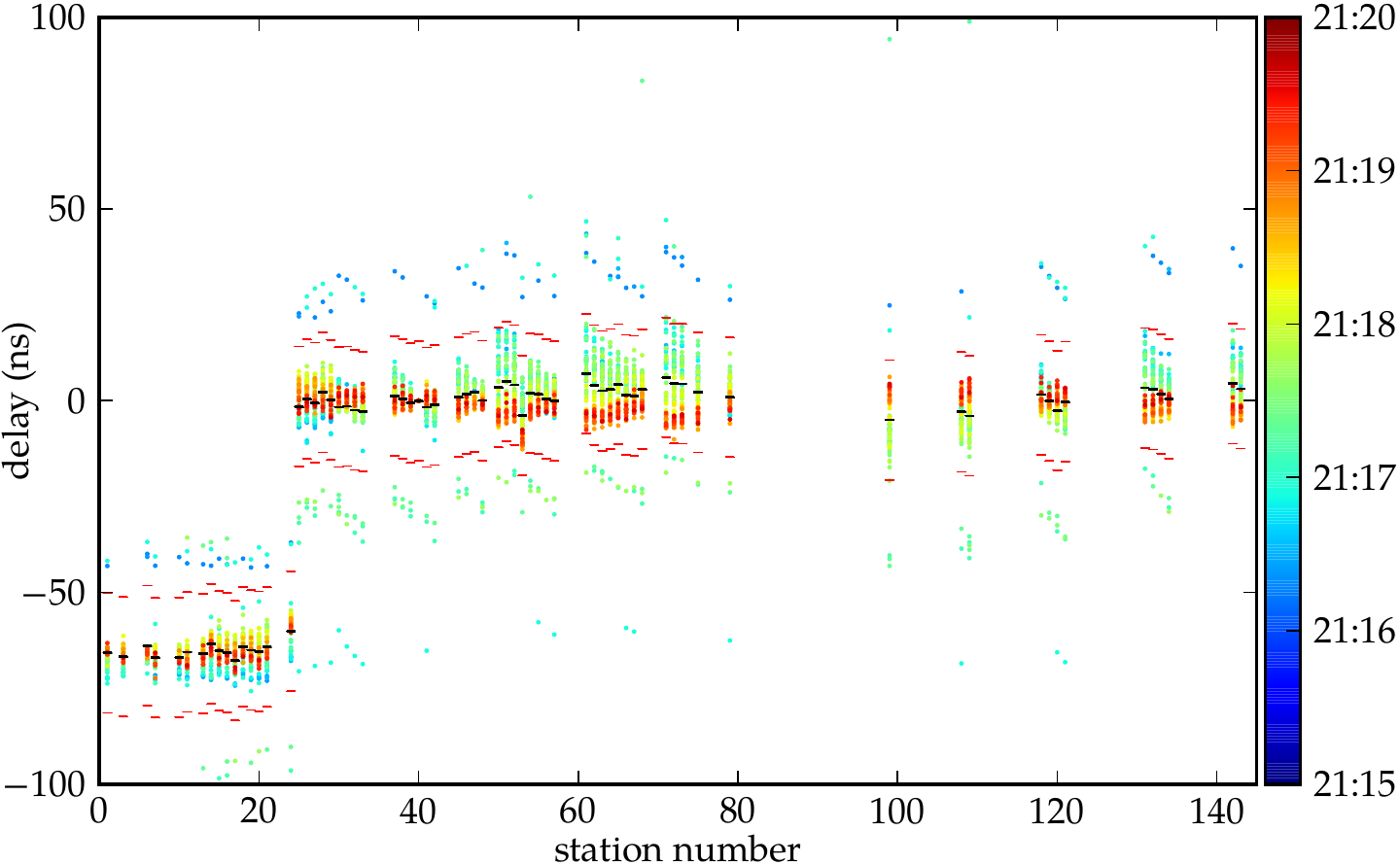}
  \caption{Same as Fig.\ \ref{fig:analysis_20140922_woB} when applying
  the airplane timing calibration after per-event clock-drift correction using the 
  beacon method. The red lines show the range of $\mathrm{mean} {}\pm{} 4 \cdot
    \mathrm{MAD}$ (median absolute deviation).
  \label{fig:analysis_oneairplane}}
\end{figure}

We first present the results of the combined beacon and airplane 
analysis inferred for the particular airplane discussed in the 
previous section. For the AERA events triggered 
by airplane pulses, we first apply the beacon timing calibration 
on an event-by-event basis, followed by the airplane timing offset 
analysis. The result is depicted in Fig.\ 
\ref{fig:analysis_oneairplane}. We can deduce several results from this 
graph.

First of all, comparing this result with the analysis of the same 
airplane events without application of the beacon correction (Fig.\ 
\ref{fig:analysis_20140922_woB}), it becomes clear 
that for stations 25 onwards, the mean timing offsets with respect to 
the reference station (number 40) indeed become 
close to zero once the beacon analysis is applied. This means that the 
beacon correction does indeed correctly compensate for GPS clock 
offsets present in the individual AERA stations.

For stations 1-24, however, there appears to be a systematic time 
offset of $\approx -65$~ns with respect to the reference station. The 
main difference between stations 1-24 and stations 25 onwards is the 
choice of antenna type \cite{AERAAntennaPaper2012}. For stations 1-24 an LPDA is used, 
whereas for stations 25 onwards we use a butterfly antenna. The two 
types of antennas have a different frequency-dependent group delay, which is
not taken into account for the beacon correction as applied for this figure. 

In principle, the different group delay can be corrected for in the data analysis on 
the basis of 4NEC2 simulations of the antenna characteristics 
\cite{AERAAntennaPaper2012, AbreuAgliettaAhn2011}. However, in the latter paper a discrepancy 
between the measured and simulated group delays is reported, in particular 
close to the horizon, i.e., for large zenith angles. This discrepancy is the likely cause of the shift between stations 
with different antenna types visible in Fig.\ \ref{fig:analysis_20140922_woB}, 
since there the simulated group delay has 
been applied in the analysis. 
Thus, the result confirms that the delays predicted by 
our antenna models probably do not correctly describe the time offsets,
which can be due to a problem in the modeling of either the LPDA, the
butterfly antenna or both of them.

While the mean time offsets are consistent among stations with the same 
type of antenna, there is very significant scatter of the offsets 
determined from individual airplane events around this mean offset. 
The scatter is obviously not random but shows two important 
characteristics. First, there is again a seemingly time-ordered distribution with 
moderate scatter (to be discussed in section \ref{sec:discussion}). Secondly, there are outliers with large time offsets beyond an obvious gap
in the distribution. The probable reason for the outliers is that some step in the airplane analysis failed. It can happen,
for example, that a neighboring peak in the enveloped airplane pulse 
is accidentally misidentified as the main peak (cf.\ Fig.\ 
\ref{fig:analysis_pulses}), especially due to the influence of noise. This leads 
to a time offset of multiples of $\approx 30$~ns and can likely explain the 
presence of the outliers with large offsets beyond the gap. We define a cut criterion 
to remove these outliers: time offsets corresponding to more than $\pm 4$~MAD (median absolute deviation)
from the mean offset of a particular station are cut away.



\subsection{Multiple airplanes}

The analysis of one airplane showed that indeed the beacon timing 
is able to determine offsets arising from GPS 
clock drifts on an event-by-event basis. However, so far we only 
showed this on a short time scale of minutes. The important question is 
now whether analyses carried out on other detected airplanes on other 
days and even in other months lead to the same consistent time 
offsets of the AERA stations.

\begin{figure}[t]
  \centering
  \includegraphics[width=0.7\textwidth]{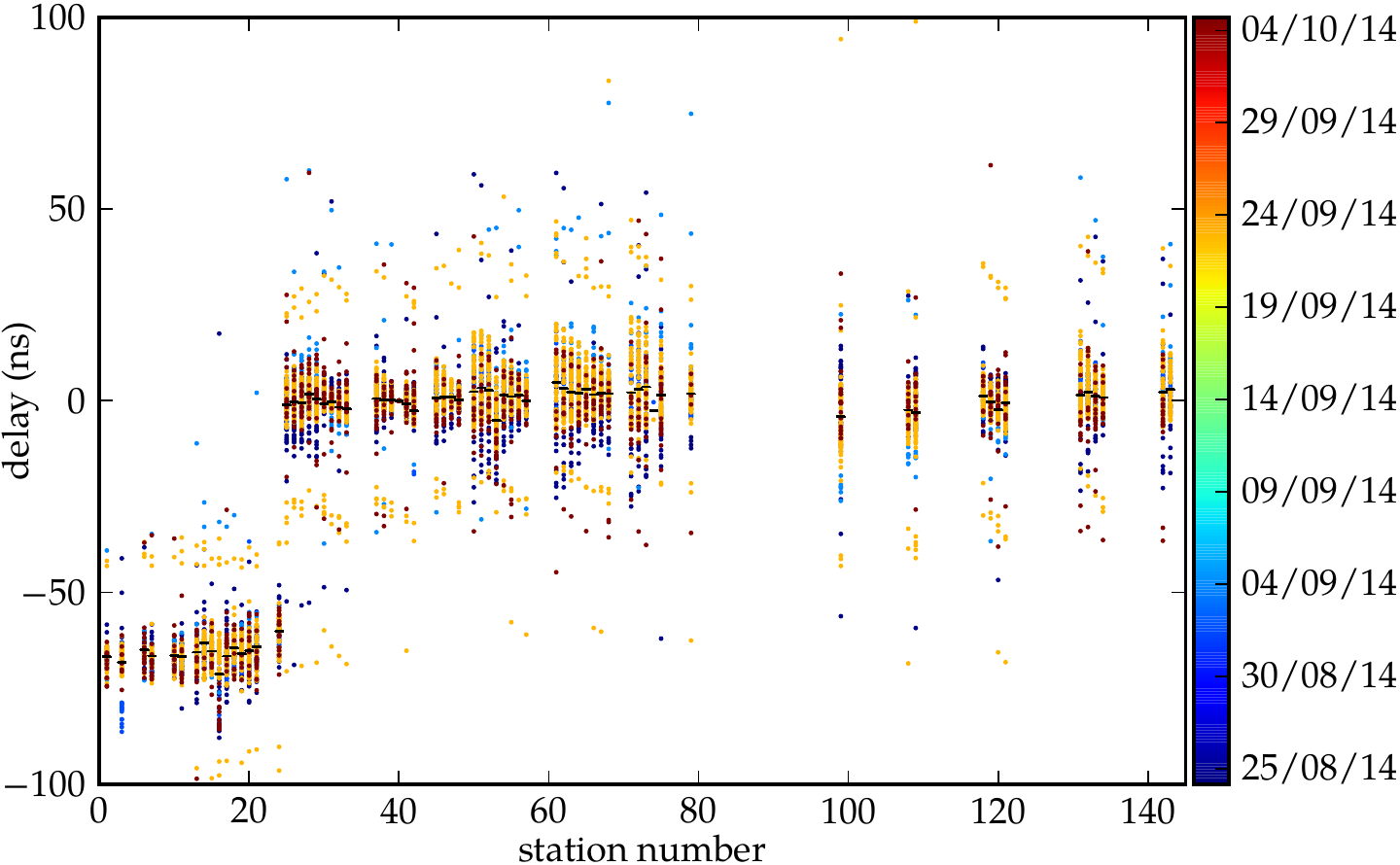}
  \caption{Same as Fig.\ \ref{fig:analysis_oneairplane} when applying 
  the combined analysis to all airplanes detected in at least 10 AERA 
  events.}
  \label{fig:analysis_2014-1004e10_wB}
\end{figure}

We apply the same combined beacon and airplane analysis as in the 
previous subsection to all airplanes that have been detected in at 
least 10 AERA events. Station 40 is chosen as the reference station in 
all cases. This results in the data shown in Fig.\ 
\ref{fig:analysis_2014-1004e10_wB}.

It becomes clear that indeed the time offsets for AERA stations 1-24 
are consistently established with a mean of $\approx -65$~ns and the 
time offsets of the stations with butterfly antennas are consistently 
very close to zero with respect to the reference station 40. The fact that 
these results have been achieved from many different airplanes over 
the course of months illustrates that both methods 
give consistent results over this long time scale and that indeed the 
beacon timing method is able to correct for true drifts arising in the 
GPS clocks of the AERA stations.

\begin{figure}
  \centering
  \includegraphics[width=0.8\textwidth]{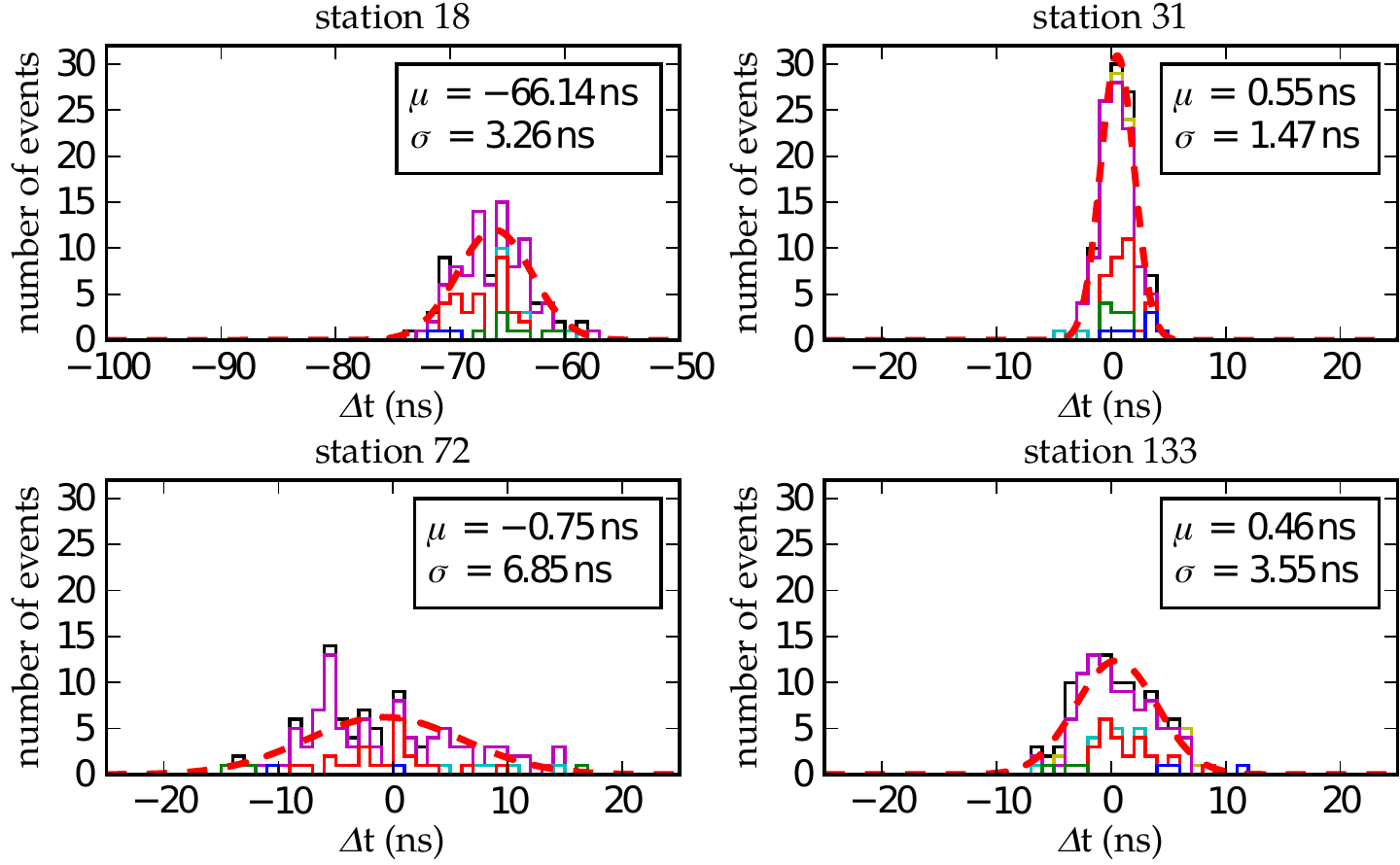}
  \caption{Time offsets determined from events of all detected 
  airplanes in AERA stations 18 (LPDA antenna),
    31, 72 and 133 (butterfly antennas). The mean time offsets $\mu$ 
    and their standard deviations
    $\sigma$ have been calculated after the MAD cut explained in section 
    \ref{sec:one_airplane}.}
  \label{fig:analysis_eventsinstations}
\end{figure}

We again cut away outliers in the resulting time offsets of
more than $\pm 4$~MAD for each individual airplane separately and show 
more quantitative results for the resulting distributions in Fig.\ 
\ref{fig:analysis_eventsinstations}. The sub-histograms with different 
colors show the resulting distributions for different airplanes. The 
mean time offsets are again consistent with the results presented for the one 
airplane in the previous subsection.

\begin{figure}[t]
  \centering
  \includegraphics[width=0.6\textwidth]{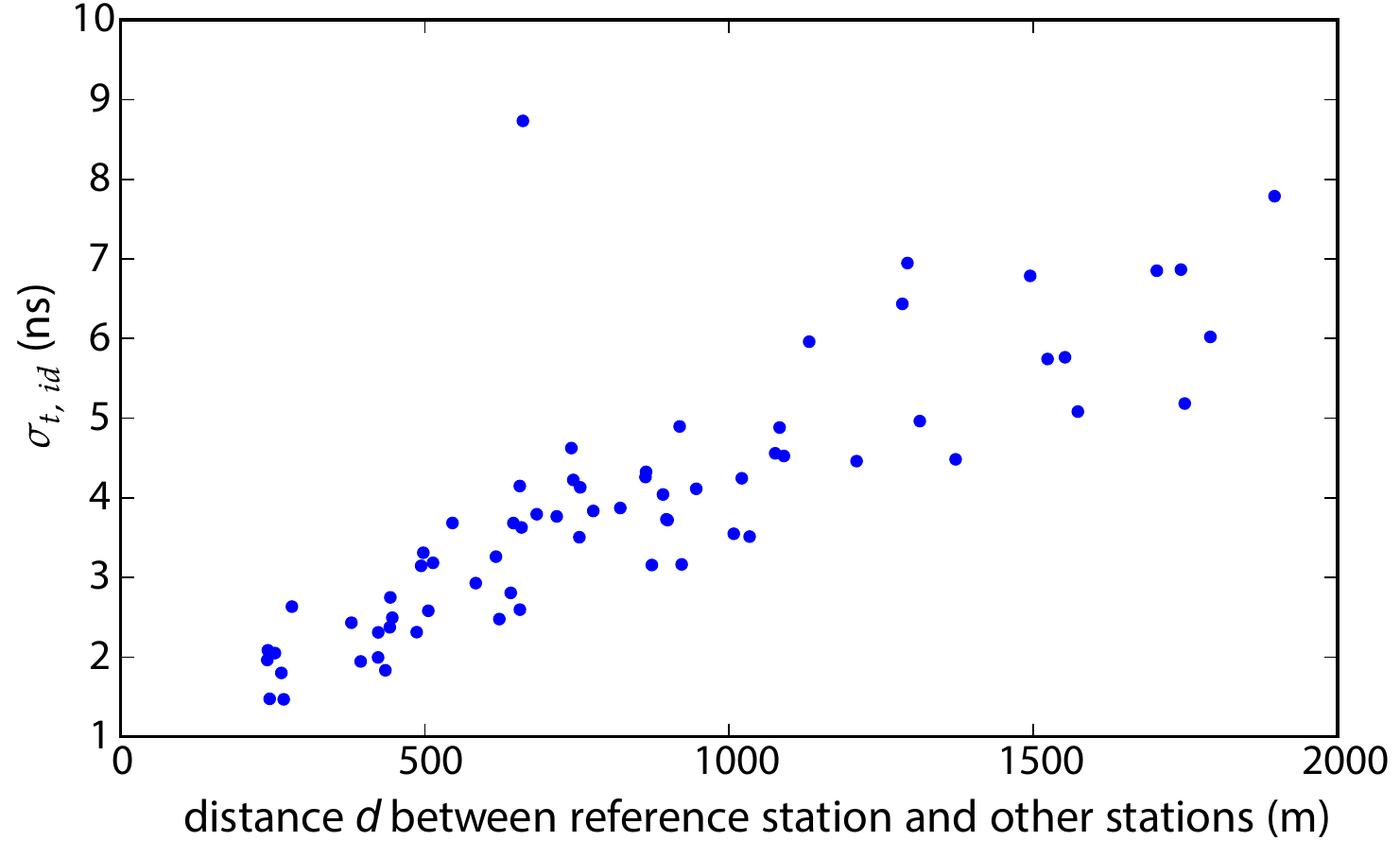}
  \caption{Standard deviation $\sigma_{t,\mathrm{id}}$ as a function of the
    distance $d$ between the reference station 40 and each individual
    AERA station $\mathrm{id}$. Each point marks a station.}
  \label{fig:analysis_sigma_distance}
\end{figure}

\begin{figure}
  \centering
  \includegraphics[width=0.7\textwidth]{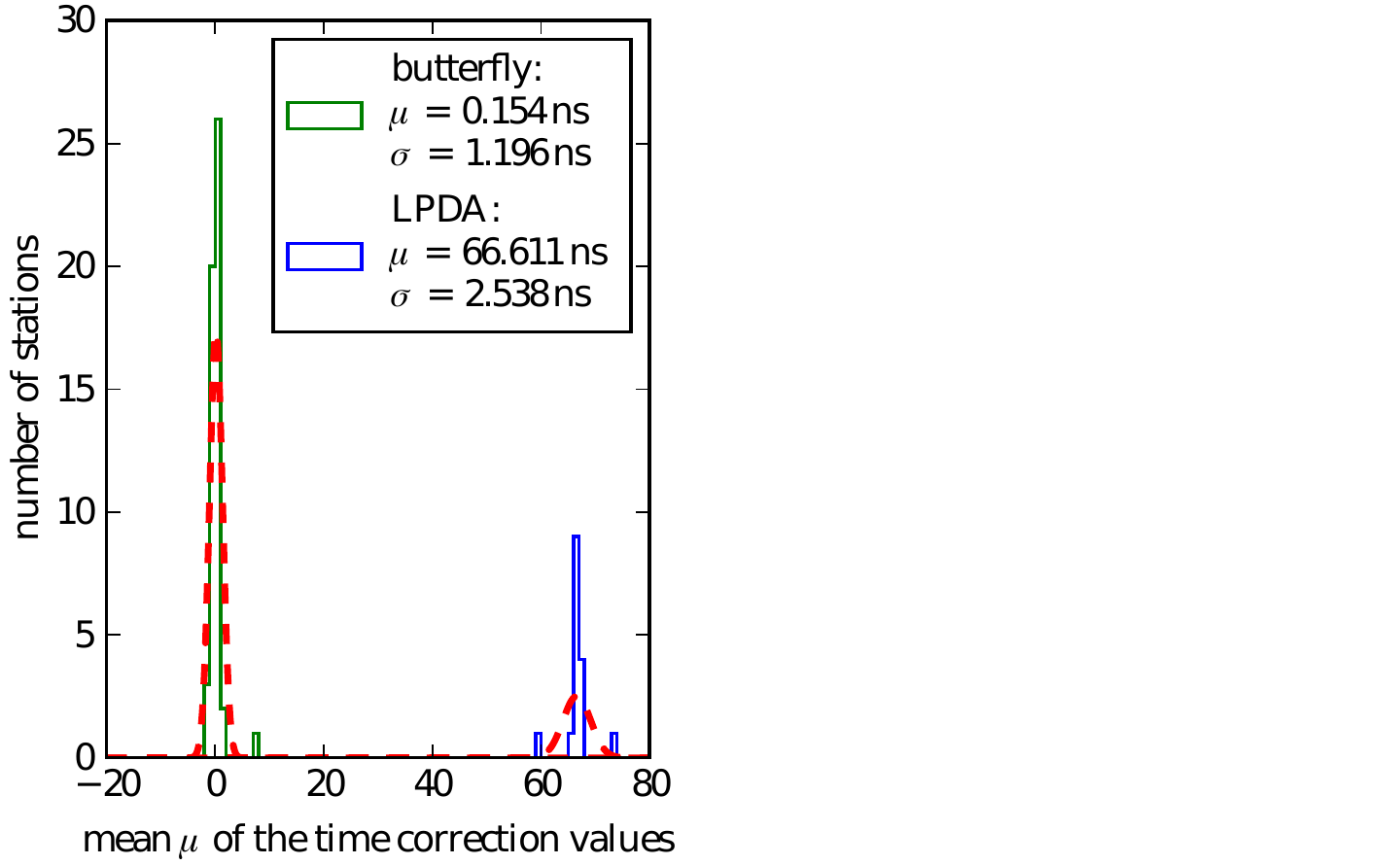}
  \caption{Histogram of the mean $\mu$ of the time correction values. 
  The standard deviations $\sigma$ of the distributions provide a measure for the average agreement between the beacon and airplane methods. 
  The values stated in the statistics box are from fitted Gaussians (dashed lines).}
  \label{fig:analysis_sigma_distance_Hist}
\end{figure}

A closer look at the distributions determined for individual AERA 
stations reveals that the widths of the resulting time-offset 
distributions are not constant but show a correlation with the 
distance of the chosen AERA station with respect to the reference station 40, see Fig.\ 
\ref{fig:analysis_sigma_distance}. This effect is expected: the 
larger the ``baseline'' over which the time offsets have to be 
determined, the more precisely the position of the airplane needs to 
be known. A more sophisticated analysis which does not rely on one 
particular station as the reference station but uses different 
reference stations for different parts of the detector array could possibly improve 
the results, at least lead to a more homogeneous width of the 
distributions among all AERA stations. We are, however, mostly 
interested in the means of the distributions, which are already 
determined well enough for our purposes with the current analysis. This
is illustrated in Fig.\ \ref{fig:analysis_sigma_distance_Hist}, which 
demonstrates that the mean time offsets determined over the course of 
several months from multiple airplane fly-overs are consistent within 
$\approx 2$~ns among AERA stations using the same antenna type. The 
small number of outliers visible in Figs.\ \ref{fig:analysis_sigma_distance} and 
\ref{fig:analysis_sigma_distance_Hist} should be investigated in 
some more detail once significantly more statistics are available.

\subsection{Reconstruction improvement}

As an example for the improvement in cosmic-ray data analysis achieved 
with the time calibration, we demonstrate the effect on the 
reconstruction of the radio wavefront of a cosmic-ray event. The 
wavefront of radio emission from cosmic-ray showers is known to be of 
hyperbolic shape \cite{LOPESWavefront2014,LOFARWavefront2014}. We 
analyze an AERA event detected simultaneously with butterfly antennas and 
LPDAs and fit a hyperbolic wavefront of the form
\begin{equation}
t = \beta \left(\sqrt{(1 + x^{2}/\gamma^{2})} - 1\right)
\end{equation}
to the measured cosmic-ray data. $\beta$ and $\gamma$ are fit 
parameters of the hyperbola, $t$ denotes the arrival time of the radio pulse with respect to 
the arrival time expected for a plane wave, and $x$ denotes the distance of
the radio detector station from the air-shower axis. The air-shower 
axis is determined by the arrival direction and the core position 
(``impact point'') of the extensive air shower. The arrival direction is well-established from the Auger surface detector 
reconstruction featuring an angular resolution of better than $1^\circ$ \cite{AugerDescriptionPaper2015}.
Thus, the arrival direction is fixed in the wavefront fit, but the core position 
reconstructed with the surface detector has significant uncertainties. 
Hence, the core position is optimized as part of 
the hyperbolic wavefront fit procedure. 

\begin{figure*}
  \centering
  \includegraphics[width=0.48\textwidth]{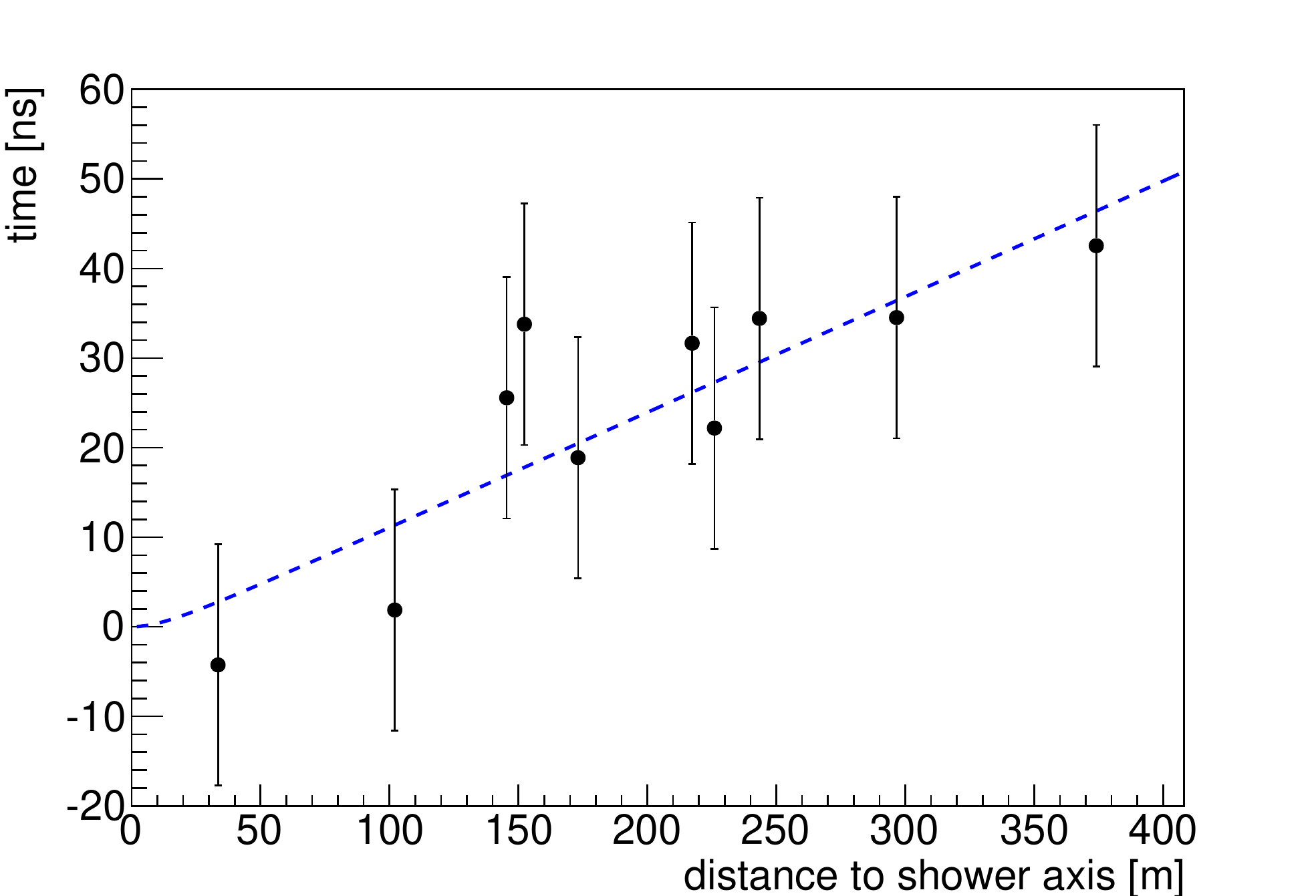}
  \includegraphics[width=0.48\textwidth]{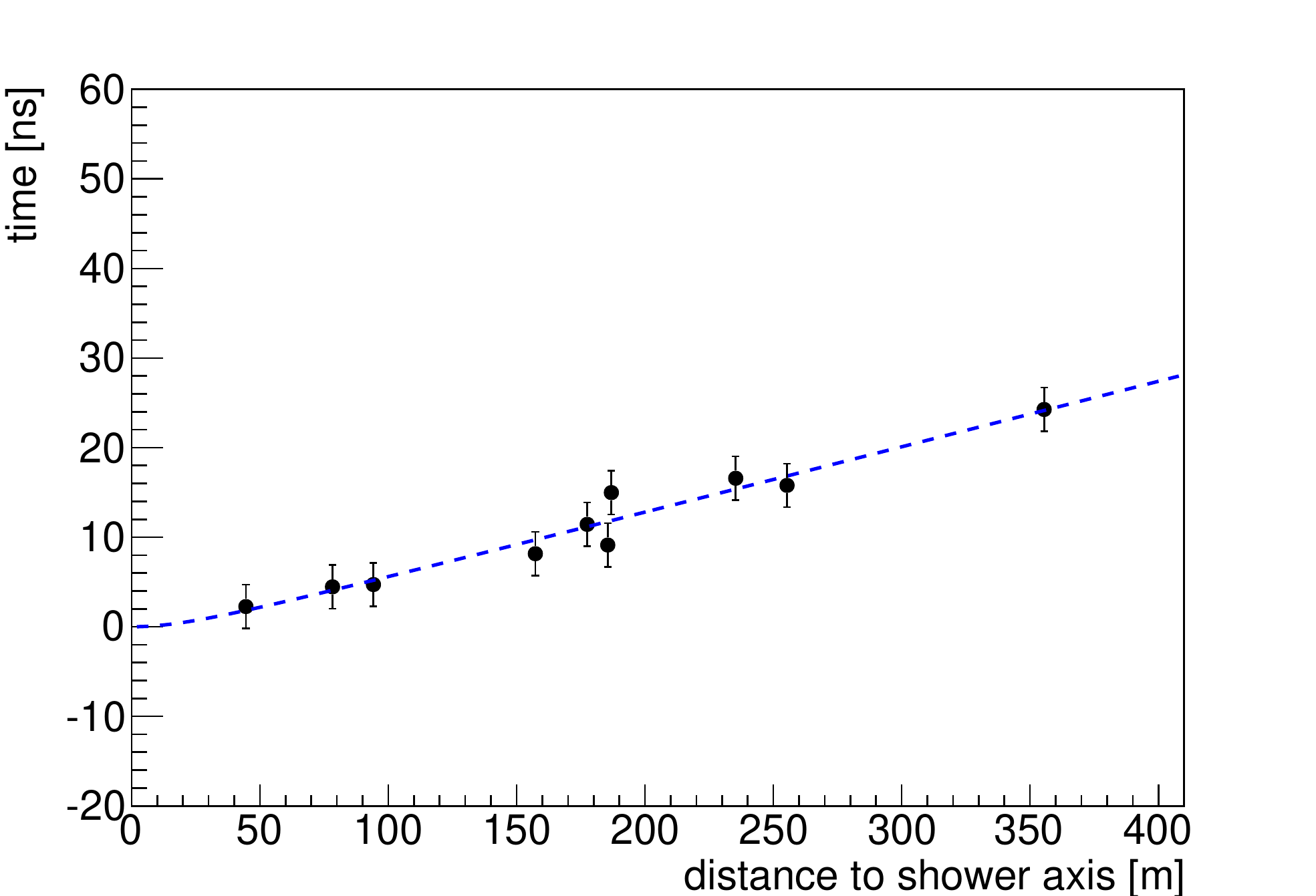}
  \caption{Improvement achieved in the reconstruction of the 
  hyperbolic radio wavefront for a cosmic-ray event measured simultaneously
  with LPDAs and butterfly antennas of AERA. The left panel shows 
  the result of a hyperbolic wavefront fit without application of 
  the time calibration. The seemingly valid fit yields a core position 
  incompatible with the one reconstructed by the Auger surface 
  detector. The right panel shows the fit result once the time 
  calibration is applied (in both panels uncertainty bars are the timing uncertainty, before and after applying the beacon method, respectively). 
  The hyperbolic wavefront fits the measured 
  pulse arrival times well and the resulting core position is in good 
  agreement with the surface-detector reconstruction.}
  \label{fig:wavefrontfit}
\end{figure*}

In Fig.\ \ref{fig:wavefrontfit} we show a comparison between a 
wavefront fit without application of the time calibration (left) and 
with application of the time calibration (right).
The improvement in the wavefront fit due to the time calibration is 
obvious. Even more importantly, however, the fit result with time calibration 
yields a core position well within the 1$\sigma$ error ellipse of the 
core position determined with the surface-detector reconstruction. The core position determined without time calibration, 
in contrast is incompatible with the surface-detector core position (it lies outside 
the 4$\sigma$ error ellipse). This example clearly illustrates the success 
and importance of the time calibration using the beacon transmitter and 
airplanes.

%


\section{Discussion} \label{sec:discussion}

The analysis presented here showed conclusively that both the beacon-timing
calibration method and the airplane-timing calibration method 
give consistent results. With this we have reached our main goal, to 
demonstrate that the beacon-timing calibration can continuously 
correct for GPS clock drifts. Also, we have established a systematic 
time offset between stations with LPDA and butterfly antennas that is 
not correctly accounted for by our current antenna models. A detailed 
verification of our antenna models for all possible directions, 
however, is very difficult with this approach as only selected arrival 
directions are available in the measurement. Activities are thus ongoing 
within AERA to cross-check the antenna models with dedicated 
calibration campaigns employing reference sources on an octocopter. This will
be important especially for air-shower events recorded simultaneously with the 
two different antenna types, since model uncertainties cancel out for those events
recorded only in stations with a unique antenna type.

The airplane-timing calibration is in principle not needed anymore in 
the future, as the beacon-timing correction has been shown to work 
reliably. For this reason we did not invest further work to improve the 
airplane analysis.  There are a number of improvements that could still be made. For 
example, the refractive index of the atmosphere is currently set to a 
constant value, resulting in a typical delay of a few ns. A 
height-dependent refractive index gradient can be applied in the 
analysis for increased precision. Tests with different choices for the 
refractive index, however, revealed that this improvement will be small compared to the 
size of the systematic effects observed.

As mentioned before, the 
choice of a fixed reference station for the whole analysis leads to 
unevenly distributed systematic uncertainties, which increase with the 
distance from the reference station. An analysis strategy using 
different reference stations for different parts of the array could 
distribute these uncertainties in a more homogeneous fashion.

The altitude of the airplanes reported in the ADS-B packets is the 
barometric height, which we interpreted as altitude above ground in our analysis, 
but in principle could be corrected for atmospheric conditions at the time of flight. 
This could yield a 
more precise airplane location and thus improved accuracy of the 
analysis. Likewise, interpolation of the airplane position between 
ADS-B data points could be improved, trying to take into account 
changes in heading and altitude occurring over the course of the 
trajectory. 

The most important issue that should be investigated further is the 
origin of the time ordering observed in the event-by-event time offsets determined 
with the airplane calibration. These indicate some direction dependence of the delays, which is 
difficult to study in detail, since the recorded airplanes cover a limited range in zenith and azimuth, only (cf.~Fig.\ \ref{fig:analysis_2014_ePLAN-0925}).
This time-ordering effect could possibly be related to a systematic offset between the reported ADS-B position 
and the position of the radio pulse transmitter onboard the aircraft. The fact that the mean 
values determined from the distributions are in such good 
agreement among detector stations with the same type of antenna gives 
us confidence that the mean values are unbiased. Nevertheless, this 
effect should be studied further, in particular when data with larger 
statistics and from more directions are available.

Assuming that the uncertainty of the ADS-B position is dominated by the 
difference between barometric and geometric height, we estimated the size of the effect 
on the basis of pressure variations in the atmospheric models used for the Auger fluorescence detectors.
Indeed the altitude difference can be on the order of several hundred 
meters in extreme cases. 
A change of altitude of this order is sufficient to change the results for individual airplanes, 
and can reduce systematic effects, e.g., the time ordering or the distance effect shown in Fig.~\ref{fig:analysis_sigma_distance}. 
However, we do not know the true airplane position. Also, analyses with varying 
altitudes revealed that an altitude correction cannot explain all effects simultaneously, and the average agreement 
between beacon and airplane analysis remains at the level of $2\,$ns. 
Consequently, the effect of the uncertainty in the airplane position 
has to be studied in more detail, and the 
dominant uncertainty in the altitude cannot be the only explanation for the systematic uncertainties of the airplane method.

An attractive option for the future could be to perform a targeted search for radio pulses from the 
International Space Station or commercial satellites. If these do emit 
radio pulses in our frequency band, they can 
be used as a regular source for timing calibration. As these sources 
are at larger distances on well-defined orbits, the achievable precision 
should be better than for airplanes, and also observation over a wider 
range of zenith and azimuth angles would be possible.

Finally, it is worth investigating the potential of using airplane 
signals also for other purposes, in particular relative amplitude 
calibration and antenna directivity studies of radio detector stations.



\section{Conclusions} \label{sec:conclusions}

The Auger Engineering Radio Array needs a relative timing accuracy 
on the order of a nanosecond to enable interferometric and wavefront analyses of the 
radio emission from extensive air showers. GPS timing alone cannot 
provide sufficient timing accuracy as there are significant drifts 
occurring in the GPS clocks. To achieve the desired timing in this 
distributed detector only connected via wireless communications, we employ 
two methods for precise time calibration. The AERA reference beacon 
transmits four continuous sine wave signals which are recorded in our 
physics data. From the relative phases of the received sine waves, GPS clock 
drifts can be corrected. Up to now it was unclear whether the drifts seen by 
the beacon represent only intrinsic GPS clock drifts or whether they could also 
be influenced by propagation and environmental effects. Therefore, we
have cross-checked the beacon method with an independent time-calibration technique. 
We can now exclude that the observed drifts are an artifact of the beacon analysis, since the 
calibration with airplane signals from different directions yields consistent results.

A low-cost software-defined radio solution continuously monitors the 
vicinity of AERA for ADS-B signals transmitted by commercial 
airplanes. These signals contain real-time position information about 
airplanes at a distance of hundreds of km. Some airplanes also emit 
pulsed radio signals in the frequency range of AERA (30-80~MHz). If an airplane approaches 
AERA, the central data acquisition is notified to let triggers from 
pulsed airplane signals pass through. An offline analysis using the 
position information from the ADS-B data and the radio pulses recorded 
in the AERA data is then able to establish the mean time offsets 
between AERA stations. A combined analysis with the beacon-timing 
correction shows that both are consistent within $\approx 2\,$ns. Consequently
each individual method is accurate to at least $2\,$ns. One of the methods, 
i.e., either the beacon or the airplane method, could well have a better accuracy than this
since the comparison then would be dominated by the less accurate 
method. In addition to this result, a previously 
unknown offset of several times $10\,$ns between AERA stations with LPDA and 
butterfly antennas was established with the airplane calibration. 

The analysis shown here establishes that the beacon-timing 
calibration works as intended and can be used for event-by-event time 
calibration of all data acquired with AERA. The accuracy is of 
$\approx 2\,$ns or better, and future tests will show if the desired accuracy of $1\,$ns is achieved already.

\acknowledgments

We would like to thank several people who supported this work directly 
or by their involvement in preparatory studies: H. Bolz, H. Bozdog, 
D. Huber, M. Konzack, R. Rink, F. Leven. We are also grateful to 
the anonymous referee for useful suggestions and comments.

The successful installation, commissioning, and operation of the Pierre Auger
Observatory would not have been possible without the strong commitment and
effort from the technical and administrative staff in Malarg\"ue. We are
very grateful to the following agencies and organizations for financial
support:

\begin{sloppypar}
Comisi\'on Nacional de Energ\'{\i}a At\'omica,
Agencia Nacional de Promoci\'on Cient\'{\i}fica y Tecnol\'ogica (ANPCyT),
Consejo Nacional de Investigaciones Cient\'{\i}ficas y T\'ecnicas (CONICET),
Gobierno de la Provincia de Mendoza,
Municipalidad de Malarg\"ue,
NDM Holdings and Valle Las Le\~nas, in gratitude for their continuing cooperation over land access,
Argentina;
the Australian Research Council (DP150101622);
Conselho Nacional de Desenvolvimento Cient\'{\i}fico e Tecnol\'ogico (CNPq), Financiadora de Estudos e Projetos (FINEP),
Funda\c{c}\~ao de Amparo \`a Pesquisa do Estado de Rio de Janeiro (FAPERJ),
S\~ao Paulo Research Foundation (FAPESP) Grants No.\ 2010/07359-6 and No.\ 1999/05404-3,
Minist\'erio de Ci\^encia e Tecnologia (MCT),
Brazil;
Grant No.\ MSMT-CR LG13007, No.\ 7AMB14AR005, and the Czech Science Foundation Grant No.\ 14-17501S,
Czech Republic;
Centre de Calcul IN2P3/CNRS, Centre National de la Recherche Scientifique (CNRS),
Conseil R\'egional Ile-de-France,
D\'epartement Physique Nucl\'eaire et Corpusculaire (PNC-IN2P3/CNRS),
D\'epartement Sciences de l'Univers (SDU-INSU/CNRS),
Institut Lagrange de Paris (ILP) Grant No.\ LABEX ANR-10-LABX-63,
within the Investissements d'Avenir Programme Grant No.\ ANR-11-IDEX-0004-02,
France;
Bundesministerium f\"ur Bildung und Forschung (BMBF),
Deutsche Forschungsgemeinschaft (DFG),
Finanzministerium Baden-W\"urttemberg,
Helmholtz Alliance for Astroparticle Physics (HAP),
Helmholtz-Gemeinschaft Deutscher Forschungszentren (HGF),
Ministerium f\"ur Wissenschaft und Forschung, Nordrhein Westfalen,
Ministerium f\"ur Wissenschaft, Forschung und Kunst, Baden-W\"urttemberg,
Germany;
Istituto Nazionale di Fisica Nucleare (INFN),
Istituto Nazionale di Astrofisica (INAF),
Ministero dell'Istruzione, dell'Universit\'a e della Ricerca (MIUR),
Gran Sasso Center for Astroparticle Physics (CFA),
CETEMPS Center of Excellence, Ministero degli Affari Esteri (MAE),
Italy;
Consejo Nacional de Ciencia y Tecnolog\'{\i}a (CONACYT),
Mexico;
Ministerie van Onderwijs, Cultuur en Wetenschap,
Nederlandse Organisatie voor Wetenschappelijk Onderzoek (NWO),
Stichting voor Fundamenteel Onderzoek der Materie (FOM),
Netherlands;
National Centre for Research and Development, Grants No.\ ERA-NET-ASPERA/01/11 and No.\ ERA-NET-ASPERA/02/11,
National Science Centre, Grants No.\ 2013/08/M/ST9/00322, No.\ 2013/08/M/ST9/00728 and No.\ HARMONIA 5 - 2013/10/M/ST9/00062,
Poland;
Portuguese national funds and FEDER funds within Programa Operacional Factores de Competitividade through Funda\c{c}\~ao para a Ci\^encia e a Tecnologia (COMPETE),
Portugal;
Romanian Authority for Scientific Research ANCS,
CNDI-UEFISCDI partnership projects Grants No.\ 20/2012 and No.\ 194/2012,
Grants No.\ 1/ASPERA2/2012 ERA-NET, No.\ PN-II-RU-PD-2011-3-0145-17 and No.\ PN-II-RU-PD-2011-3-0062,
the Minister of National Education,
Programme Space Technology and Advanced Research (STAR), Grant No.\ 83/2013,
Romania;
Slovenian Research Agency,
Slovenia;
Comunidad de Madrid,
FEDER funds,
Ministerio de Educaci\'on y Ciencia,
Xunta de Galicia,
European Community 7th Framework Program, Grant No.\ FP7-PEOPLE-2012-IEF-328826,
Spain;
Science and Technology Facilities Council,
United Kingdom;
Department of Energy, Contracts No.\ DE-AC02-07CH11359, No.\ DE-FR02-04ER41300, No.\ DE-FG02-99ER41107 and No.\ DE-SC0011689,
National Science Foundation, Grant No.\ 0450696,
The Grainger Foundation,
USA;
NAFOSTED,
Vietnam;
Marie Curie-IRSES/EPLANET,
European Particle Physics Latin American Network,
European Union 7th Framework Program, Grant No.\ PIRSES-2009-GA-246806;
and
UNESCO.
\end{sloppypar}

\clearpage

\section*{Appendix: Detected Airplanes}
\label{app:analysis_airplaneevents}

\begin{table*}[h]
  \centering
  \caption{Airplanes detected with AERA in the time from June 20 
  to October 5, 2014. Only detections fulfilling the following quality 
  cuts are listed: detection in at least 10 
  detector stations with a signal-to-noise ratio of at least 15 and a 
  reconstructed radius of curvature of the spherical wavefront between 
  7.5~km and 99~km as well as a reconstructed zenith angle less than 
  80$^{\circ}$. We do not know, why the number of events varies strongly, i.e.,
  why a few airplanes emit significantly more detectable pulses than others.}
  \vspace{0.5cm}
  \label{tab:analysis_airplaneevents}
    \begin{tabular}{@{}ccccc@{}}
      \toprule
      UTC date & events & Mode-S call sign & airplane type & airline\\ \midrule
      20/06/14, 16:26 & 3  & C05843 & Boeing 767-35H  & Air Canada \\
      22/07/14, 03:26 & 1  & E48987 & Boeing 737-8EH  & GOL Transportes \\
      22/07/14, 03:34 & 6  & E47FE0 & Airbus A320-232 & TAM Linhas Aereas \\
      28/07/14, 15:43 & 1  & E06541 & Boeing 737-81D  & Aerolineas Argentinas \\
      31/07/14, 06:29 & 1  & E80320 & Airbus A319-111 & Sky Airline \\
      31/07/14, 11:28 & 4  & E48C03 & Boeing 737-8HX  & GOL Transportes \\
      31/07/14, 18:20 & 1  & E48987 & Boeing 737-8EH  & GOL Transportes \\
      03/08/14, 11:27 & 1  & E48986 & Boeing 737-8EH  & GOL Transportes \\
      24/08/14, 03:36 & 16 & E48854 & Boeing 737-8EH  & GOL Transportes \\
      24/08/14, 11:30 & 17 & E48854 & Boeing 737-8EH  & GOL Transportes \\
      01/09/14, 11:23 & 32 & E48854 & Boeing 737-8EH  & GOL Transportes \\
      03/09/14, 21:21 & 11 & E48C04 & Boeing 737-8HX  & GOL Transportes \\
      08/09/14, 21:33 & 9  & E48C03 & Boeing 737-8HX  & GOL Transportes \\
      08/09/14, 22:07 & 4  & E80411 & Airbus A320-233 & LAN Airlines \\
      10/09/14, 01:58 & 1  & E80413 & Airbus A320-233 & LAN Airlines \\
      13/09/14, 13:02 & 3  & E80414 & Airbus A320-233 & LAN Airlines \\ 
      22/09/14, 21:16 & 115& E48854 & Boeing 737-8EH  & GOL Transportes \\
      28/09/14, 21:27 & 1  & E48C04 & Boeing 737-8HX  & GOL Transportes \\
      02/10/14, 14:27 & 1  & 0C208E & Boeing 737-8V3  & Copa Airlines   \\
      04/10/14, 11:28 & 14 & E48985 & Boeing 737-8EH  & GOL Transportes \\
      04/10/14, 17:13 & 33 & E48853 & Boeing 737-8EH  & GOL Transportes \\ \bottomrule
    \end{tabular}
\end{table*}


\end{document}